\documentclass[10pt,twocolumn,twoside]{IEEEtran}

\usepackage{amsmath,amssymb}

\DeclareMathOperator*{\subjectto}{subject\;to}

\usepackage{graphicx}

\usepackage{subcaption}

\usepackage{cite}

\newtheorem{theorem}{Theorem}
\newtheorem{definition}{Definition}
\newtheorem{lemma}{Lemma}

\newtheorem{corollary}{Corollary}
\newtheorem{remark}{Remark}

\usepackage{xcolor}

\usepackage{bbm}

\usepackage[normalem]{ulem} 

\newcommand{\Expect}[0]{\mathbb{E}}
\newcommand{\xSize}[0]{n_{\boldsymbol{X}}}
\newcommand{\ySize}[0]{n_{\boldsymbol{Y}}}

\title{Quadratic Privacy-Signaling Games and the MMSE Information Bottleneck Problem\\ for Gaussian Sources}

\author{
Ertan~Kaz{\i}kl{\i},
Sinan~Gezici,~\IEEEmembership{Senior~Member,~IEEE,}
and~Serdar~Y\"uksel,~\IEEEmembership{Member,~IEEE}
\thanks{
Part of this work was presented at the 2020 IEEE International Symposium on Information Theory (ISIT), June 21-26, 2020,  Los Angeles, California, USA \cite{ISITVersion}.

E. Kaz{\i}kl{\i}, and S. Y\"uksel are with the Department of Mathematics and Statistics, Queen's University, K7L 3N6, Kingston, Ontario, Canada. Emails: ertan.kazikli@queensu.ca and yuksel@mast.queensu.ca. 

S. Gezici is with the Department of Electrical and Electronics Engineering, Bilkent University, 06800, Ankara, Turkey, Email: gezici@ee.bilkent.edu.tr.}}

\begin{document}
\maketitle

\begin{abstract}
We investigate a privacy-signaling game problem in which a sender with privacy concerns observes a pair of correlated random vectors which are modeled as jointly Gaussian. The sender aims to hide one of these random vectors and convey the other one whereas the objective of the receiver is to accurately estimate both of the random vectors. We analyze these conflicting objectives in a game theoretic framework with quadratic costs where depending on the commitment conditions (of the sender), we consider Nash or Stackelberg (Bayesian persuasion) equilibria. We show that a payoff dominant Nash equilibrium among all admissible policies is attained by a set of explicitly characterized linear policies. We also show that a payoff dominant Nash equilibrium coincides with a Stackelberg equilibrium. We formulate the information bottleneck problem within our Stackelberg framework under the mean squared error distortion criterion where the information bottleneck setup has a further restriction that only one of the random variables is observed at the sender. We show that this MMSE Gaussian Information Bottleneck Problem admits a linear solution which is explicitly characterized in the paper. We provide explicit conditions on when the optimal solutions, or equilibrium solutions in the Nash setup, are informative or noninformative. 
\end{abstract}


\section{Introduction and System Model}\label{sec:intro}

Various applications such as social networks, networked control, smart grid and crowd sensing benefit from data collected from decision makers. In these applications, users share information with a service provider which wishes to improve the quality of service by utilizing information gathered from the users. The users as well are interested in enhanced service quality as they benefit from the service while at the same time they wish to retain a certain level of privacy. The privacy objective arises from the fact that the information they wish to convey to the service provider may be correlated with certain private information they want to protect. For instance, in smart grid applications, power usage information shared by the users with the service provider may disclose some information related to users such as their habits and behaviors \cite{McDaniel2009,PrivacySmartMeteringSurvey}. For that reason, privacy is a major challenge in smart grid applications and this is a current research topic in numerous studies (see \cite{PrivacySmartMeteringSurvey,HanACC2016,CompetiviePrivacy2011,YaoAllerton2013} and references therein). In addition, the problem of preserving privacy while maintaining reasonable system performance appears in various contexts \cite{CrowdSensingWirelessComm2015,PrivacyCrowdSensingIoT2016,Yamamoto1983,EstEffUndPriConstr2019,WangIT2019,PadakandlaIT2020,DiazIT2020,UtilityvsPrivacy2013,UtilityvsPrivacy2018,SreekumarISIT2020,Calmon2012,LuAutomatica2020,DiffPrivateFilteringTAC2014,PrivacyConsensusTAC2017,Nekouei2019,WangTCNS2017}.

In this manuscript, we consider the following communication scenario between a sender and a receiver motivated by the aforementioned applications. There is a pair of sources at the sender and the perspective of the sender is such that one source needs to be protected and the other source needs to be conveyed. As opposed to the sender, the receiver desires to accurately estimate both sources with the aim of acquiring as much information as possible. Under this setting, we investigate the interactions between the sender and the receiver whose objectives are different from each other due to the privacy concerns of the sender. 

\begin{figure}
\centering
\null\hfill
\begin{subfigure}[t]{0.48\textwidth}
\centering
\includegraphics[scale=0.75]{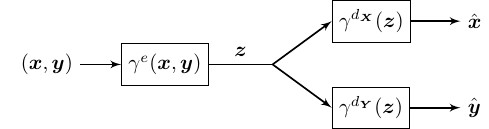}
\caption{Privacy-signaling game.}
\label{fig:blockDiagram}
\end{subfigure}
\hfill\vspace{5pt}
\begin{subfigure}[t]{0.48\textwidth}
\centering
\includegraphics[scale=0.75]{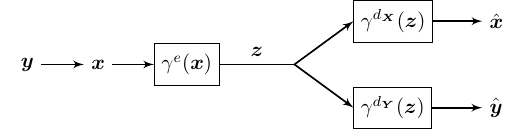}
\caption{Information bottleneck.}
\label{fig:blockDiagramIB}
\end{subfigure}
\hfill\null
\caption{Information flow under the quadratic privacy-signaling game and the information bottleneck problems.}
\label{figure:allScheme}
\end{figure}

Consider an information transmission scenario in which a sender encodes a pair of correlated random vectors $\boldsymbol{X}$ and $\boldsymbol{Y}$ into $\boldsymbol{Z}$ using an encoding function denoted by $\boldsymbol{z}=\gamma^e(\boldsymbol{x},\boldsymbol{y})$ and a receiver wants to decode both of the random vectors based on its observation $\boldsymbol{Z}=\boldsymbol{z}$. We denote the size of the random vectors $\boldsymbol{X}$ and $\boldsymbol{Y}$ by $\xSize$ and $\ySize$, respectively. In this communication scenario, the sender wishes to convey information contained in $\boldsymbol{Y}$ whereas it views $\boldsymbol{X}$ as a private random variable that needs to be hidden from the receiver. The aim of the receiver is to accurately estimate both of the random variables given its observation $\boldsymbol{Z}=\boldsymbol{z}$. Let the decoders for estimating $\boldsymbol{X}$ and $\boldsymbol{Y}$ at the receiver be denoted by $\gamma^{d_{\boldsymbol{X}}}(\boldsymbol{z})$ and $\gamma^{d_{\boldsymbol{Y}}}(\boldsymbol{z})$, respectively. Fig.~\ref{fig:blockDiagram} illustrates the considered information transmission scenario.

We model the random variables $\boldsymbol{X}$ and $\boldsymbol{Y}$ as jointly Gaussian random vectors. Let $
\begin{bmatrix}
\boldsymbol{X} \\ \boldsymbol{Y}
\end{bmatrix}$ be a zero mean Gaussian random vector with a positive definite covariance matrix $\Sigma\triangleq \begin{bmatrix}
\Sigma_{\boldsymbol{X}} & \Sigma_{\boldsymbol{X}\boldsymbol{Y}} \\
\Sigma_{\boldsymbol{Y}\boldsymbol{X}} & \Sigma_{\boldsymbol{Y}}
\end{bmatrix}$. The random variables $\boldsymbol{X}$ and $\boldsymbol{Y}$ are not independent of each other, i.e., $\Sigma_{\boldsymbol{X}\boldsymbol{Y}}\neq \boldsymbol{0}$. It is assumed that the joint distribution of $\boldsymbol{X}$ and $\boldsymbol{Y}$ is common knowledge, i.e., both players know $\Sigma_{\boldsymbol{X}}$, $\Sigma_{\boldsymbol{Y}}$ and $\Sigma_{\boldsymbol{X}\boldsymbol{Y}}$. Since $\boldsymbol{X}$ and $\boldsymbol{Y}$ are correlated, transmitting $\boldsymbol{Y}$ directly discloses information related to the private random variable $\boldsymbol{X}$. In other words, the objectives of hiding $\boldsymbol{X}$ and conveying $\boldsymbol{Y}$ are conflicting. These conflicting objectives at the sender are modeled via the following objective function:
\begin{align}
&J^e(\gamma^e,\gamma^{d_{\boldsymbol{X}}},\gamma^{d_{\boldsymbol{Y}}}) \nonumber \\
&\hphantom{J^e(\gamma^e}
=\Expect[\lVert \boldsymbol{Y}-\gamma^{d_{\boldsymbol{Y}}}(\boldsymbol{Z})\rVert^2]
-\delta\,\Expect[\lVert \boldsymbol{X}-\gamma^{d_{\boldsymbol{X}}}(\boldsymbol{Z})\rVert^2],\label{eq:objEnc}
\end{align}
which is to be minimized, where $\delta$ is a positive design parameter that determines the level of desired privacy in terms of hiding $\boldsymbol{X}$. On the other hand, the receiver aims to extract both of the random variables. Thus, the receiver wishes to minimize the following objective function:
\begin{align}\label{eq:objDec}
&J^d(\gamma^e,\gamma^{d_{\boldsymbol{X}}},\gamma^{d_{\boldsymbol{Y}}}) \nonumber \\
&\hphantom{J^e(\gamma^e}
=\Expect[\lVert \boldsymbol{Y}-\gamma^{d_{\boldsymbol{Y}}}(\boldsymbol{Z})\rVert^2]
+\Expect[\lVert \boldsymbol{X}-\gamma^{d_{\boldsymbol{X}}}(\boldsymbol{Z})\rVert^2].
\end{align}
In \eqref{eq:objDec}, the mean squared errors corresponding to random variables $\boldsymbol{X}$ and $\boldsymbol{Y}$ are incorporated into the objective function with equal weights since taking different weights does not alter the problem. In this work, we investigate the Nash equilibrium and the Stackelberg equilibrium for the described strategic information transmission scenario in which the objectives of the sender and the receiver are as defined above.

The game dynamics for the Nash equilibrium are as follows: The players choose their strategies simultaneously. These chosen strategies are referred to as a Nash equilibrium if no player gains by unilaterally deviating from its strategy. In other words, neither the sender nor the receiver have any incentive to unilaterally change their strategies when they operate at a Nash equilibrium. Suppose that the set of possible strategies for the encoder is denoted by $\Gamma^e$, i.e., $\gamma^e\in\Gamma^e$, and those for the decoders of each random variable are denoted by $\Gamma^{d_{\boldsymbol{X}}}$ and $\Gamma^{d_{\boldsymbol{Y}}}$, i.e., $\gamma^{d_{\boldsymbol{X}}}\in\Gamma^{d_{\boldsymbol{X}}}$ and $\gamma^{d_{\boldsymbol{Y}}}\in\Gamma^{d_{\boldsymbol{Y}}}$. A set of policies $\gamma^{e,*}$, $\gamma^{d_{\boldsymbol{X}},*}$ and $\gamma^{d_{\boldsymbol{Y}},*}$ forms a Nash equilibrium if \cite{TBasarBook}
\begin{align}
J^e(\gamma^{e,*},\gamma^{d_{\boldsymbol{X}},*},\gamma^{d_{\boldsymbol{Y}},*})
\leq
J^e(\gamma^e,\gamma^{d_{\boldsymbol{X}},*},\gamma^{d_{\boldsymbol{Y}},*}) 
\end{align}
for all $\gamma^{e}\in \Gamma^e$ and
\begin{align}
J^d(\gamma^{e,*},\gamma^{d_{\boldsymbol{X}},*},\gamma^{d_{\boldsymbol{Y}},*})
\leq
J^d(\gamma^{e,*},\gamma^{d_{\boldsymbol{X}}},\gamma^{d_{\boldsymbol{Y}}})
\end{align}
for all $\gamma^{d_{\boldsymbol{X}}}\in \Gamma^{d_{\boldsymbol{X}}}$ and $\gamma^{d_{\boldsymbol{Y}}}\in \Gamma^{d_{\boldsymbol{Y}}}$.

\begin{remark}
Under the Nash equilibrium concept, there is no commitment assumption for the players. This may be appropriate for scenarios where the players do not trust the announcements of each other or do not have access to policy announcements. For instance, a user (sender) making sensor measurements in a crowd sensing application may encounter a tradeoff between utility of providing useful information to a data aggregator (receiver) and protecting its privacy. We may consider a setting where the sender has the ability to reconfigure its policy. In this case, the sender wishes to deviate from a certain announced policy if it knows that such deviation leads to a better privacy protection given the receiver's announcement. Thus, the receiver does not trust the policy announcement of the sender. On the other hand, the sender may also think that the receiver's announcement is not trustworthy. This happens for instance when the receiver discloses collected information from individuals to third parties which do not comply with the receiver's commitment. In addition, the sender may wish to guard itself against data breaches at the legitimate receiver. In order to model such scenarios where both the sender and the receiver do not have any commitment regarding their policies, a Nash theoretic game model can be used. Although they do not commit to a certain policy, if they are in equilibrium, then they do not wish to deviate unilaterally.
\end{remark}

On the other hand, the Stackelberg equilibrium involves a sequential game play in the sense that first the sender and then the receiver act (this setup is commonly referred to as the {\it Bayesian persuasion} problem in the economics literature \cite{kamenica2011bayesian}). The sender chooses and announces its strategy and then the receiver acts upon learning the strategy of the sender. Here, the sender commits to employ this announced strategy. The receiver employs an optimal response to the announced strategy of the sender. A set of policies $\gamma^{e,*}$, $\gamma^{d_{\boldsymbol{X}},*}$ and $\gamma^{d_{\boldsymbol{Y}},*}$ forms a Stackelberg equilibrium if \cite{TBasarBook}
\begin{align}
J^e(\gamma^{e,*},\gamma^{d_{\boldsymbol{X}},*}(\gamma^{e,*}),&\gamma^{d_{\boldsymbol{Y}},*}(\gamma^{e,*}))\nonumber\\
&\leq
J^e(\gamma^{e},\gamma^{d_{\boldsymbol{X}},*}(\gamma^{e}),\gamma^{d_{\boldsymbol{Y}},*}(\gamma^{e}))
\end{align}
for all $\gamma^{e}\in \Gamma^e$, where $\gamma^{d_{\boldsymbol{X}},*}(\gamma^{e})$ and $\gamma^{d_{\boldsymbol{Y}},*}(\gamma^{e})$ are such that
\begin{align}
J^d(\gamma^{e},\gamma^{d_{\boldsymbol{X}},*}(\gamma^{e}),&\gamma^{d_{\boldsymbol{Y}},*}(\gamma^{e}))\nonumber\\
&\leq
J^d(\gamma^{e},\gamma^{d_{\boldsymbol{X}}}(\gamma^{e}),\gamma^{d_{\boldsymbol{Y}}}(\gamma^{e}))
\end{align}
for all $\gamma^{d_{\boldsymbol{X}}}\in \Gamma^{d_{\boldsymbol{X}}}$ and $\gamma^{d_{\boldsymbol{Y}}}\in \Gamma^{d_{\boldsymbol{Y}}}$.

\begin{remark}
It is important to emphasize that under the Stackelberg equilibrium concept, there is a commitment assumption for the sender and the sender cannot backtrack its commitment. This setup can be appropriate for scenarios where an information provider publicly shares information given its observations. For instance, consider a medical research setting. Researchers wish to publicly reveal data they obtained as a result of medical research so that other researchers can benefit from this data. However, as this data may contain sensitive information related to participants of the study, the researchers need to take privacy into account while publishing their research data. In this case, the researchers employ a privacy-preserving data revelation scheme so that the privacy of the participants is protected. On the other hand, in order for other researchers to make sense of this revealed research data, they need to know what type of privacy-preserving mechanism is employed in the design. Therefore, the researchers performing the study publicly reveal the specification of such mechanism they used. This corresponds to a scenario with sender commitment, as in the Stackelberg setup considered in this manuscript. 
\end{remark}

We will also consider an instance of the problem above as the {\it MMSE Gaussian Information Bottleneck Problem}. The difference with the setup above is that the sender only has access to $\boldsymbol{X}$, which is the message it intends to hide while revealing as much information on $\boldsymbol{Y}$ as possible. This is depicted in Fig.~\ref{fig:blockDiagramIB}. The classical information bottleneck problem \cite{tishby1999information} considers the mutual information as the cost criterion where the aim is to compress an observed random variable while preserving information related to an unobserved correlated random variable. Note that both the privacy and the compression objectives aim at removing the corresponding information from the revealed message. Motivated by this resemblance, we consider the information bottleneck problem in our game theoretic context. Details are provided in Section \ref{sec:IB}.

\subsection{Literature Review}

For signaling games under the Nash equilibrium concept, Crawford and Sobel in their foundational paper \cite{CrawfordSobel} investigate a communication scenario between a sender and a receiver where sender's cost contains a bias term leading to misaligned objectives. They obtain the interesting result that under some technical conditions the sender needs to quantize the information it sends at a Nash equilibrium. To put it differently, the misalignment in the objectives results in information hiding through quantization of the transmitted message. In contrast to Crawford and Sobel, the Bayesian persuasion problem\cite{kamenica2011bayesian} investigates signaling scenarios under the Stackelberg equilibrium concept rather than the Nash equilibrium concept. 

In the context of the Bayesian persuasion problem, an important related work \cite{Tamura2018} considers a multidimensional signaling scenario under a Stackelberg game setup where the sender employs a general quadratic cost structure. An upper bound on the performance of the sender is obtained via formulating a semidefinite program. For jointly Gaussian sources, a linear policy that achieves this upper bound is characterized which shows the optimality of linear policies for such a Stackelberg game. We use this characterization in some of our results rather prominently.

Recently, the strategic information transmission (SIT) problem has attracted attention also in the communication and control theory literature \cite{Ekyol2017ProcIEEE,Saritas2017QuadraticEquilibria,EstStrategicSensorsFarokhi2017,SubjectiveBiasesSITBasar2018,treust2017persuasion,treust2018persuasion,SaritasAutomatica2020,SaritasISIT2019,SayinAutomatica2019,sayin2019optimal,MultiCheapArxiv}. For instance, the work in \cite{Saritas2017QuadraticEquilibria} considers quadratic costs with a bias term appearing in sender's cost and investigates both scalar and multidimensional source settings. An interesting observation from \cite{Saritas2017QuadraticEquilibria} is the existence of a linear Nash equilibrium which is in contrast to the quantized nature of the equilibrium in Crawford and Sobel. In \cite{EstStrategicSensorsFarokhi2017}, the misalignment in the objectives is due to a bias term which is modeled as a random variable. The authors consider the Stackelberg equilibrium concept and focus on affine policies. In \cite{SubjectiveBiasesSITBasar2018}, a communication scenario between prospect theoretic agents whose cost functions are distorted by subjective biases is investigated using the Stackelberg equilibrium concept.

In the literature, several studies consider the SIT problem in which the sender takes privacy of certain information into account by employing a suitable privacy measure, under either the Nash or Stackelberg criteria \cite{Farokhi2015QuadraticGames,Akyol2015PrivacyProcessing,Farokhi2016PrivacyCommunication,EAkyolITW2015}. In these studies, a common theme is to model private and nonprivate random variables as jointly Gaussian random variables. In \cite{Farokhi2015QuadraticGames}, a communication scenario between a sender and a receiver is investigated using the Stackelberg equilibrium concept in which an additional side information is assumed to be available at the receiver. The estimation errors are measured using quadratic costs and a family of Stackelberg equilibria is characterized under an \textit{a priori} affine policy assumption. In contrast, here, we do not restrict the policies to be affine \textit{a priori} and we consider a setting with no side information. We investigate Nash equilibria as well and show that a payoff dominant equilibrium is attained by linear policies. We also show that these linear policies at the payoff dominant Nash equilibrium lead to a Stackelberg equilibrium even when the encoding policy is not restricted to be linear. The work in \cite{Akyol2015PrivacyProcessing} also investigates a Stackelberg game where the utility measure for the nonprivate random variable is quadratic and the privacy measure is entropy based. Both noiseless and noisy communication scenarios are considered and essentially unique linear encoding and decoding policies that form a Stackelberg equilibrium are characterized. In \cite{Farokhi2016PrivacyCommunication}, a Nash game is studied where the privacy measure is based on mutual information and the utility measure for the nonprivate random variable is quadratic. In \cite{Farokhi2016PrivacyCommunication}, apart from the previously described Gaussian scenario, another scenario in which private and nonprivate data are treated as discrete random variables is considered.

The tradeoff between utility and privacy appears also in various other contexts \cite{DifPrivacy2008,AlgoDifPrivacy2014,
Yamamoto1983,EstEffUndPriConstr2019,WangIT2019,PadakandlaIT2020,DiazIT2020,UtilityvsPrivacy2013,UtilityvsPrivacy2018,SreekumarISIT2020,Calmon2012,LuAutomatica2020,Calmon2015,IBtoPrivacyFunnel2014,DiffPrivateFilteringTAC2014,HanTAC2017,PrivacyConsensusTAC2017,Nekouei2019,WangTCNS2017}. One line of related work is the differential privacy literature where the main problem of interest is to protect private information on publicly available databases \cite{DifPrivacy2008,AlgoDifPrivacy2014}. The notion of differential privacy ensures that private information provided by an individual to a database is not compromised by a third party, e.g., a data analyst, who retrieves information from this database. In this context, an interesting result from \cite{DiffPrivateFilteringTAC2014} is the application of the Laplacian or Gaussian perturbations to guarantee differential privacy. For a comprehensive treatment of differential privacy on such problems as filtering and estimation, please see \cite{jeromeBook}. Another line of work is the privacy funnel problem \cite{Calmon2015} where it is desired to convey as much information as possible related to an observed random variable while trying to leak as low information as possible related to an unobserved private random variable. It should be noted that in the privacy funnel problem, only the nonprivate random variable is observed at the sender whereas in our framework, we assume that both the nonprivate and private random variables are observed at the sender. Another related work is \cite{EstEffUndPriConstr2019} where the tradeoff between utility and privacy is investigated through formulating constrained optimization problems that consider settings with a discrete random variable and a continuous random variable. The continuous random variable case focuses on Gaussian perturbations applied to the nonprivate random variable to protect private information.

As noted, a further related problem is the information bottleneck problem \cite{tishby1999information,TishbyITW2015,HarremoesISIT2007,Gilad-Bachrach2003,Dhillon2003,VeraISIT2018,VeraIT2019,GoldfeldJSAIT2020,HsuISIT2018,ZaidiEntropy2020} which also has connections with the privacy funnel problem \cite{IBtoPrivacyFunnel2014}. In the information bottleneck technique \cite{tishby1999information}, the aim is to compress an observed random variable while trying to preserve information related to another correlated random variable which is not observed. It is important to note that the information bottleneck technique is closely related to an earlier seminal work \cite{WitsenhausenIT1975} which considers a similar constrained optimization problem by employing conditional entropy to asses the performance. The information bottleneck problem specializing to Gaussian sources is investigated in \cite{chechik2005IB} where the random variables of interest are jointly Gaussian random vectors. The compression objective in the information bottleneck problem can also be viewed as a privacy objective as in our framework in the sense that the corresponding information is desired to be removed from the revealed message. In the information bottleneck problem, the costs involve mutual information and only one of the random variables is received at the sender whereas in our framework the costs include mean squared error terms and both of the random variables are observed at the sender. In order to provide an estimation theoretic perspective on the information bottleneck problem, we formulate a similar problem where we use mean squared error terms for the costs as in our original setting and we show that there are operational and consequential differences when the encoder is allowed to use both of the hidden variables.

\subsection{Contributions of the Manuscript}
The main aim of this manuscript is to provide both Nash and Stackelberg equilibria analyses for the considered privacy-signaling game problem. In game theory, since the solution concept involves an equilibrium (Nash, Stackelberg, and refinements), one cannot talk about an optimal equilibrium in general. Nonetheless, as a main contribution of our work, we establish and compute an equilibrium, which is desirable among all, for both of the players. The main contributions of this manuscript can be summarized as follows:
\begin{itemize}
\item In the literature, a Nash equilibrium analysis of the privacy game problem, in which both the privacy and the utility (for the nonprivate random variable) are measured via the mean squared error cost, has not been available. In this manuscript, we consider this problem for the first time in the literature to our knowledge. More importantly, we show that a payoff dominant Nash equilibrium is attained by linear policies in Theorem~\ref{thm:NashVector}. These equilibria are the most desirable equilibria for both of the players among any set of policies. We show that the characterized linear payoff dominant Nash equilibria coincide with the Stackelberg equilibria in Theorem~\ref{thm:StackelbergVector}. It should be emphasized that these (Stackelberg) equilibria are obtained without an \textit{a priori} affine policy restriction for the players. In other words, if we consider the optimization problem that the encoder needs to solve while obtaining the Stackelberg equilibria, these linear policies are the optimal solution among any sets of policies. 
\item We introduce an MMSE Gaussian information bottleneck problem, which is a modification of the classical information bottleneck problem that has been considered under mutual information criteria. By viewing this as an instance of the privacy-signaling game under the Stackelberg formulation, we show that the solution to the MMSE Gaussian information bottleneck problem is attained by a set of explicitly characterized linear policies in Theorem~\ref{thm:IB}. Namely, even when the policies are allowed to be nonlinear, a set of linear policies arises as the optimal solution.
\item We extend our results for scalar sources to the additive Gaussian noise channel setting. Under this setting, it is shown that a payoff dominant Nash equilibrium is attained by linear policies in Theorem~\ref{thm:NashNoisy}. This theorem also establishes that the characterized linear Nash equilibrium is unique among the affine class. We also show that the payoff dominant Nash equilibrium coincides with the Stackelberg equilibrium in Theorem~\ref{thm:StackelbergNoisy}. In addition, the characterized linear Stackelberg equilibrium is unique among any set of policies. We also establish the existence of nonlinear Nash and Stackelberg equilibria considering a discrete channel setting in which the encoding function is restricted to take discrete values in Theorem~\ref{thm:NashDiscrete} and Theorem~\ref{thm:StackelbergDiscrete}, respectively.
\end{itemize}

\subsection{Organization of the Manuscript}

The remainder of the manuscript is organized as follows. Section~\ref{sec:Nash} and Section~\ref{sec:Stackelberg} provide, respectively, the Nash and Stackelberg equilibria analyses for the considered privacy-signaling game. Section~\ref{sec:IB} investigates the information bottleneck problem as an instance of our proposed framework. Section~\ref{sec:noisy} and Section~\ref{sec:discreteChannel} extend the results for scalar sources to the Gaussian noise channel and discrete channel, respectively. Section~\ref{sec:nume} provides numerical examples, and  Section~\ref{sec:conc} concludes the manuscript with some final remarks.

\section{Nash Equilibria} \label{sec:Nash}

In this section, we characterize linear Nash equilibria of the considered privacy-signaling game. More importantly, we show that special cases of these equilibria lead to payoff dominant Nash equilibria. These payoff dominant Nash equilibria are the most desirable equilibria for both of the players (among all coding/decoding policies, including those that are nonlinear) in a sense that is made explicit in the following definition.

\begin{definition}
A Nash equilibrium that is not Pareto dominated\footnote{A set of policies $\gamma^e(\cdot,\cdot)$, $\gamma^{d_{\boldsymbol{X}}}(\cdot)$ and $\gamma^{d_{\boldsymbol{Y}}}(\cdot)$ Pareto dominates another set of policies $\tilde{\gamma}^e(\cdot,\cdot)$, $\tilde{\gamma}^{d_{\boldsymbol{X}}}(\cdot)$ and $\tilde{\gamma}^{d_{\boldsymbol{Y}}}(\cdot)$ if 
$J^e(\gamma^e,\gamma^{d_{\boldsymbol{X}}},\gamma^{d_{\boldsymbol{Y}}})\leq J^e(\tilde{\gamma}^e,\tilde{\gamma}^{d_{\boldsymbol{X}}},\tilde{\gamma}^{d_{\boldsymbol{Y}}})$, $J^d(\gamma^e,\gamma^{d_{\boldsymbol{X}}},\gamma^{d_{\boldsymbol{Y}}})\leq J^d(\tilde{\gamma}^e,\tilde{\gamma}^{d_{\boldsymbol{X}}},\tilde{\gamma}^{d_{\boldsymbol{Y}}})$ and at least one of these inequalities is strict.} by any other Nash equilibrium of the game is said to be a payoff dominant Nash equilibrium \cite{HarsanyiSeltenBook1988} . 
\end{definition}

In order to characterize linear Nash equilibria, we propose an equivalent formulation by applying an invertible linear transformation of variables from Tamura \cite{Tamura2018}. We note that \cite{Tamura2018} considers a general multidimensional signaling setup under quadratic costs and characterizes a set of linear policies that forms a Stackelberg equilibrium for jointly Gaussian sources. We use this characterization for our special case of privacy-signaling game scenario to formulate an equivalent problem and this approach facilitates our Nash equilibrium analysis.

\begin{theorem}\label{thm:NashVector}
\begin{enumerate}
\item[(i)] There exist informative\footnote{We refer to an equilibrium as noninformative if the sender does not convey information related to both of the random variables at this equilibrium and this is equivalent to what is known as a {\it babbling equilibrium} in the signaling games literature. In the converse case, the equilibrium is referred to as informative.} linear Nash equilibria with an encoding policy
\begin{align}\label{eq:NashEnc}
\boldsymbol{z} = 
\begin{bmatrix}
\alpha_1 \boldsymbol{q}_1 & \cdots & \alpha_{\ySize} \boldsymbol{q}_{\ySize}
\end{bmatrix}^T
\Sigma^{-1/2} 
\begin{bmatrix}
\boldsymbol{x} \\ \boldsymbol{y}
\end{bmatrix}
\end{align}
for scalars $\{\alpha_i\}_{i=1}^{\ySize}$ with at least one of these scalars being nonzero\footnote{The case when $\alpha_i=0$ for all $i=1,\dots,\ySize$ leads to a noninformative Nash equilibrium.}
where $\{\boldsymbol{q}_i\}_{i=1}^n$ are normalized eigenvectors of $W=\Sigma^{1/2} \mathrm{diag}(-\delta I,I) \Sigma^{1/2}$ with $n\triangleq \xSize+\ySize$ and these eigenvectors are arranged in such a way that the corresponding eigenvalues $\{\lambda_i\}_{i=1}^n$ satisfy $\lambda_i>0$ for $i=1,\dots,\ySize$ and $\lambda_i<0$ for $i=\ySize+1,\dots,n$. The corresponding decoding policy is given by
\begin{align}\label{eq:NashDec}
\begin{bmatrix}
\gamma^{d_{\boldsymbol{X}}}(\boldsymbol{z})\\ \gamma^{d_{\boldsymbol{Y}}}(\boldsymbol{z})
\end{bmatrix}
= \Sigma^{1/2} 
\begin{bmatrix}
\beta_1\boldsymbol{q}_1 & \cdots & \beta_{\ySize}\boldsymbol{q}_{\ySize} 
\end{bmatrix}
\boldsymbol{z}
\end{align}
where $\beta_i = 1/\alpha_i$ if $\alpha_i\neq 0$ and $\beta_i=0$ otherwise for $i=1,\dots,\ySize$.These Nash equilibria exist for any set of scalars $\{\alpha_i\}_{i=1}^{\ySize}$. When the indices with $\alpha_i\neq 0$ are the same for two sets of scalars, they lead to the same performance values, i.e., the resulting equilibria with these sets of scalars are informationally equivalent.
\item[(ii)] These informative equilibria are payoff dominant Nash equilibria if $\alpha_i\neq 0$ for all $i=1,\dots,\ySize$.
\item [(iii)] In addition to informative equilibria, there always exist noninformative Nash equilibria with the transmitted message being independent of the sources, e.g.,$\gamma^e(\boldsymbol{x},\boldsymbol{y})=C$ for some constant $C$, and with the decoding policies $\gamma^{d_{\boldsymbol{X}}}(\boldsymbol{z})=\boldsymbol{0}$ and $\gamma^{d_{\boldsymbol{Y}}}(\boldsymbol{z})=\boldsymbol{0}$.
\end{enumerate}
\end{theorem}

\begin{IEEEproof}
We apply a linear transformation of variables that gives an orthonormal coordinate system and then show that in this transformed coordinate system the sender wishes to convey some of the coordinates and to hide the remaining coordinates. The advantage of this transformation is that as these coordinates are orthogonal to each other, conveying one coordinate does not give information related to other coordinates. 

Since $W=\Sigma^{1/2}\mathrm{diag}(-\delta I,I)\Sigma^{1/2}$ is symmetric, we can decompose it as $W=Q\Lambda Q^T$ for orthonormal $Q$ and diagonal $\Lambda$. We denote the columns of $Q$ by $\{\boldsymbol{q}_i\}_{i=1}^n$ and the diagonal elements of $\Lambda$ by $\{\lambda_i\}_{i=1}^n$. By Sylvester's law of inertia, $W$ and $\mathrm{diag}(-\delta I,I)$ have the same number of positive and negative eigenvalues \cite[p.~282]{horn2013matrixAnalysis}. Therefore, $W$ has $\ySize$ positive and $\xSize$ negative eigenvalues. We sort these eigenvalues $\{\lambda_i\}_{i=1}^n$ in such a way that the first $\ySize$ of them are positive and the remaining ones are negative. For notational convenience, we define 
\begin{align}\label{eq:defS}
\boldsymbol{S} \triangleq
\begin{bmatrix}
\boldsymbol{X}\\ \boldsymbol{Y}
\end{bmatrix},\quad 
\gamma^{d_{\boldsymbol{S}}}(\boldsymbol{z}) \triangleq 
\begin{bmatrix}
\gamma^{d_{\boldsymbol{X}}}(\boldsymbol{z})\\ \gamma^{d_{\boldsymbol{Y}}}(\boldsymbol{z})
\end{bmatrix}.
\end{align}
Now, we make the following transformation of variables:
\begin{align}\label{eq:defT}
\boldsymbol{T}\triangleq Q^T\Sigma^{-1/2}
\boldsymbol{S},
\end{align}
where $\Sigma^{-1/2}$ is well-defined since $\Sigma$ is assumed to be positive definite. We note that 
\begin{align*}
\Expect[\boldsymbol{T}\boldsymbol{T}^T] 
&= 
\Expect[Q^T\Sigma^{-1/2}\boldsymbol{S}\boldsymbol{S}^T\Sigma^{-1/2}Q] \\
&= 
Q^T\Sigma^{-1/2}\Expect[\boldsymbol{S}\boldsymbol{S}^T]\Sigma^{-1/2}Q \\
&=
Q^T\Sigma^{-1/2}\Sigma\Sigma^{-1/2}Q = I.
\end{align*}
Thus, each components of $\boldsymbol{T}$ are independent and identically distributed zero-mean Gaussian random variables each with a unit variance. 

Next, we propose an equivalent problem under this linear transformation of variables. The encoder consists of two consecutive mappings, one of which is fixed as above and the other one can arbitrarily be chosen by the sender. In other words, there is a linear mapping from $(\boldsymbol{x},\boldsymbol{y})$ to $\boldsymbol{t}$ fixed as \eqref{eq:defT} and then an encoding function $\boldsymbol{z}=\tilde{\gamma}^e(\boldsymbol{t})$. At the receiver side, we also consider a similar setting. The observation at the receiver is mapped into $\tilde{\boldsymbol{t}}$ via $\gamma^{d_{\boldsymbol{T}}}(\boldsymbol{z})$, which can arbitrarily be selected by the receiver. Then, these auxiliary variables are mapped into estimates of $\boldsymbol{x}$ and $\boldsymbol{y}$ as follows:
\begin{align}\label{eq:defSEstimate}
\gamma^{d_{\boldsymbol{S}}}(\boldsymbol{z})
= \Sigma^{1/2}Q \gamma^{d_{\boldsymbol{T}}}(\boldsymbol{z}).
\end{align}
Fig.~\ref{fig:equivalentBlock} provides an illustration for the equivalent formulation. The aim is to characterize $\tilde{\gamma}^e(\boldsymbol{t})$ as well as $\gamma^{d_{\boldsymbol{T}}}(\boldsymbol{z})$ at a Nash equilibrium. Since the proposed transformation is invertible, the problem in the transform domain is equivalent to the original problem. 

\begin{figure}
\centering
\includegraphics[width=\linewidth]{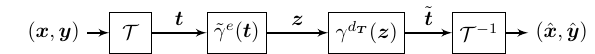}
\caption{Equivalent formulation where $\mathcal{T}$ denotes the linear transformation specified in \eqref{eq:defT}.}
\label{fig:equivalentBlock}
\end{figure}

Next, we express the objective function of each player in terms of the random variables in the transformed coordinate system. The objective function of the sender can be written as
\begin{align}
J^e(\tilde{\gamma}^e,\gamma^{d_{\boldsymbol{T}}})
&= 
\Expect[
(\boldsymbol{S}-\gamma^{d_{\boldsymbol{S}}}(\boldsymbol{Z}))^T
\mathrm{diag}(-\delta I,I) 
\nonumber \\
& \hphantom{=\Expect[(}
(\boldsymbol{S}-\gamma^{d_{\boldsymbol{S}}}(\boldsymbol{Z}))
]\nonumber \\
&= 
\Expect[
(\boldsymbol{T}-\gamma^{d_{\boldsymbol{T}}}(\boldsymbol{Z}))^T Q^T \Sigma^{1/2}
\mathrm{diag}(-\delta I,I) 
\nonumber \\
& \hphantom{=\Expect[(}
\Sigma^{1/2}Q(\boldsymbol{T}-\gamma^{d_{\boldsymbol{T}}}(\boldsymbol{Z}))
]\nonumber \\
&= 
\Expect[
(\boldsymbol{T}-\gamma^{d_{\boldsymbol{T}}}(\boldsymbol{Z}))^T 
\Lambda
(\boldsymbol{T}-\gamma^{d_{\boldsymbol{T}}}(\boldsymbol{Z}))
]\nonumber \\
&=\sum_{i=1}^n \lambda_i 
\Expect[
(T_i-\gamma^{d_{T_i}}(\boldsymbol{Z}))^2
] \label{eq:JeEqui},
\end{align}
where the first equation is obtained from \eqref{eq:objEnc} and \eqref{eq:defS}, the second equation is based on \eqref{eq:defT} and \eqref{eq:defSEstimate}, and $\{\lambda_i\}_{i=1}^n$ are the eigenvalues of $W$ which satisfy $\lambda_i>0$ for $i=1,\dots,\ySize$ and $\lambda_i<0$ for $i=\ySize+1,\dots,n$. Similarly, if we express the objective function of the receiver in terms of the random variables in the introduced coordinate system, we get 
\begin{align}
J^d(\tilde{\gamma}^e,\gamma^{d_{\boldsymbol{T}}})
& = \Expect[ 
(\boldsymbol{S}-\gamma^{d_{\boldsymbol{S}}}(\boldsymbol{Z}))^T
(\boldsymbol{S}-\gamma^{d_{\boldsymbol{S}}}(\boldsymbol{Z}))
]\nonumber \\
&=\Expect[
(\boldsymbol{T}-\gamma^{d_{\boldsymbol{T}}}(\boldsymbol{Z}))^T
K
(\boldsymbol{T}-\gamma^{d_{\boldsymbol{T}}}(\boldsymbol{Z})) 
] \label{eq:JdEqui}
\end{align}
where $K\triangleq Q^T\Sigma Q$ is a positive definite matrix since $\Sigma$ is positive definite. In the proposed equivalent problem, the optimal $\gamma^{d_{\boldsymbol{T}}}(\boldsymbol{z})$ for a given encoding policy $\tilde{\gamma}^e(\boldsymbol{t})$ is the minimum mean squared error estimator of the corresponding random variable. The proof of this statement is standard but we present it in Lemma~\ref{lem:MMSEOptimal} of Appendix~\ref{sec:supporting} for completeness.

In the equivalent formulation, the objective function of the sender is expressed as a weighted sum of mean squared error terms corresponding to independent random variables where there are both positive and negative weights. We partition the transformed random vector as
\begin{align}\label{eq:defUandV}
\boldsymbol{T} \triangleq 
\begin{bmatrix}
\boldsymbol{U} \\
\boldsymbol{V}
\end{bmatrix}
\end{align}
where $\boldsymbol{U} \in \mathbb{R}^{\ySize}$ and $\boldsymbol{V} \in \mathbb{R}^{\xSize}$ such that the positive and negative coefficients in \eqref{eq:JeEqui} correspond to $\boldsymbol{U}$ and $\boldsymbol{V}$, respectively. The receiver still employs the minimum mean squared error estimators for the corresponding random variables for a given encoding policy. Next, we use the equivalent formulation to characterize the Nash equilibria. Due to structure of sender's cost in \eqref{eq:JeEqui} considering the equivalent formulation, it follows that the sender does not convey any information related to $\boldsymbol{V}$, which is specified in \eqref{eq:defUandV}, at a Nash equilibrium. We present this auxiliary result in Lemma~\ref{lem:DoNotSendVNash} of Appendix~\ref{sec:supporting}. Hence, the transmitter is restricted to send information related to $\boldsymbol{U}$ at a Nash equilibrium. This implies that the objective function of the sender at a Nash equilibrium reduces to 
\begin{align}
J^e(\tilde{\gamma}^e,\gamma^{d_{\boldsymbol{T}}}) = \sum_{i=1}^{\ySize} \lambda_i \Expect[(T_i-\gamma^{d_{T_i}}(\boldsymbol{Z}))^2]
\end{align}
where $\{1,\dots,\ySize\}$ are the only indices with $\lambda_i> 0$ in $\{\lambda_i\}_{i=1}^n$. Since the receiver wishes to extract all the random variables in this transformed coordinate system including the ones with these indices, the receiver also shares the objective of minimizing these mean squared error terms. As a result, conveying all or a subset of the random variables in $\boldsymbol{U}$ yields a Nash equilibrium and this gives the linear policies stated in \eqref{eq:NashEnc}. In addition, conveying all of these random variables yields the minimum attainable performance at a Nash equilibrium for both of the players. Thus, if $\alpha_i\neq 0$ for all $i=1,\dots,\ySize$ in \eqref{eq:NashEnc}, the corresponding linear Nash equilibria are payoff dominant Nash equilibria. Namely, there does not exist any other Nash equilibrium which Pareto dominates these characterized equilibria. 
\end{IEEEproof}

\begin{remark}
It is interesting to contrast the result of Theorem~\ref{thm:NashVector} with the signaling game setups in the literature where the misaligned cost structure arises from biased nature of the sender as opposed to privacy concerns of the sender. Notably, Crawford and Sobel \cite{CrawfordSobel} introduce a signaling game setup where the costs are misaligned due to a bias term. The communication setup is similar to our setting in this section in the sense that the transmitted message is perfectly observed by the receiver and the sender does not have a power constraint. Under this setting, \cite{CrawfordSobel} establishes the quantized nature of Nash equilibria for scalar sources supported on $[0,1]$ under certain assumptions regarding the objectives, and \cite{Saritas2017QuadraticEquilibria} generalizes this result to arbitrary source distributions under quadratic criteria. In contrast to this setting with a biased sender, if there is a privacy concerned sender, then a Nash equilibrium is attained by linear policies regardless of source dimensions, as shown in Theorem~\ref{thm:NashVector}. In fact, depending on whether the source is scalar or vector valued, there may exist linear informative Nash equilibria for the signaling game setup with a biased sender. In particular, in \cite{MultiCheapArxiv}, we extend Crawford and Sobel's formulation to a multidimensional source setting under quadratic cost criteria with a biased sender and show that for independent and identically distributed sources and an arbitrary bias vector, there always exist linear informative Nash equilibria (only) when the source distribution is Gaussian. 
\end{remark}

Theorem~\ref{thm:NashVector} characterizes linear Nash equilibria in which there is communication between the transmitter and the receiver. Hence, the considered game always admits informative linear Nash equilibria regardless of the system parameters. More importantly, special cases of these linear equilibria leads to payoff dominant Nash equilibria, which are the most desirable equilibria for both of the players among all coding/decoding policies.

Next, we specialize to the case of scalar sources with the aim to provide a more explicit characterization of the payoff dominant Nash equilibria in this case. In particular, $X$ and $Y$ are assumed to be zero-mean jointly Gaussian with variances $\sigma_X^2$ and $\sigma_Y^2$, respectively and a nonzero correlation $\rho$, i.e., $\Sigma = \begin{bmatrix}
\sigma_X^2 & \rho \\ \rho & \sigma_Y^2
\end{bmatrix}$. We present the following as a corollary of Theorem~\ref{thm:NashVector}.

\begin{corollary}\label{cor:NashScalar}
For scalar sources $X$ and $Y$, there exist informative linear Nash equilibria with an encoding policy $\gamma^e(x,y)=A x + B y$ which satisfies
\begin{align}\label{eq:BARatio}
\frac{B}{A}
=
- \frac{\delta\sigma_X^2+\sigma_Y^2}{2\delta\rho }
- \frac{\sqrt{(\delta\sigma_X^2+\sigma_Y^2)^2-4\delta\rho^2}}{2\delta\rho}
\end{align}
and decoding policies
\begin{align}
&\gamma^{d_X}(z) =
\left(
\frac
{A\sigma_X^2+B\rho}
{A^2\sigma_X^2+B^2\sigma_Y^2+ 2AB\rho }
\right)
z, \label{eq:decXScalar}
\\
&\gamma^{d_Y}(z) =
\left(
\frac
{A\rho + B\sigma_Y^2}
{A^2\sigma_X^2+B^2\sigma_Y^2+ 2AB\rho }
\right)
z.\label{eq:decYScalar}
\end{align}
These equilibria are payoff dominant Nash equilibria. In addition, these equilibria are unique among linear policies.
\end{corollary}
\begin{IEEEproof}
In order to characterize the Nash equilibria, we need to find the eigenvalues and eigenvectors of $W = \Sigma^{1/2} \mathrm{diag}(-\delta,1) \Sigma^{1/2}$. We note that $W$ has the same eigenvalues as $D \triangleq \Sigma\, \mathrm{diag}(-\delta,1)$ and these can be computed as
\begin{align}
\lambda_1 &= 
\frac{-\delta\sigma_X^2+\sigma_Y^2}{2}
+ \frac{\sqrt{(\delta\sigma_X^2+\sigma_Y^2)^2-4\delta\rho^2}}{2},
\label{eq:defLambda1} \\
\lambda_2 &= 
\frac{-\delta\sigma_X^2+\sigma_Y^2}{2}
-\frac{\sqrt{(\delta\sigma_X^2+\sigma_Y^2)^2-4\delta\rho^2}}{2}.
\label{eq:defLambda2}
\end{align} 
where $\lambda_1>0$ and $\lambda_2 <0 $. Thus, the sender is restricted to transmit 
$
u = \boldsymbol{q}_1^T \Sigma^{-1/2} \begin{bmatrix}
x \\ y
\end{bmatrix}\label{eq:defUScalar}
$
where $\boldsymbol{q}_1$ is the normalized eigenvector of $W$ associated with the eigenvalue $\lambda_1$ computed in \eqref{eq:defLambda1}. It is seen that $\boldsymbol{q}_1^T \Sigma^{-1/2}$ is a left eigenvector of $D$ associated with its eigenvalue $\lambda_1$. By expressing this left eigenvector, an encoding policy which satisfies \eqref{eq:BARatio} is obtained. Then, the conditional expectation formula for jointly Gaussian distributions can be employed to obtain \eqref{eq:decXScalar} and \eqref{eq:decYScalar} \cite[p. 155]{PoorBook}. As a result, these characterized policies lead to payoff dominant Nash equilibria. Moreover, the only possible linear equilibria are attained by transmitting a scaled version of $u$.
\end{IEEEproof}

\begin{remark}
Note that when $\delta\rightarrow 0^+$, the encoding policy satisfies $A/B\rightarrow 0$ as can be deduced from \eqref{eq:BARatio}. Therefore, $\delta\rightarrow 0^+$ implies that the encoder transmits directly $Y$ at a Nash equilibrium. When $\delta=0$, the sender also transmits $Y$ directly at a Nash equilibrium since it does not have any privacy concern in this case. Hence, the Nash equilibrium specified in Corollary~\ref{cor:NashScalar} as $\delta\rightarrow 0^+$ coincides with the Nash equilibrium when $\delta=0$.
\end{remark}

\begin{remark}\label{rem:highPrivacy}
It is seen that the ratio of $B$ and $A$ converges to $-\sigma_x^2/\rho$ as $\delta\rightarrow\infty$, which can be verified from \eqref{eq:BARatio}. This shows that in the high privacy regime, the encoder makes the revealed information $Z$ and the private random variable $X$ essentially uncorrelated.
\end{remark}

\section{Stackelberg Equilibria} \label{sec:Stackelberg}

In this section, we characterize the Stackelberg equilibria of the considered quadratic privacy-signaling game. Our main result is that the payoff dominant Nash equilibria characterized in the previous section are also Stackelberg equilibria. It is important to note that the set of possible encoding strategies, i.e., $\Gamma^e$, is not restricted to be linear. If the sender performs an optimization of its objective function by anticipating the best response of the receiver, these linear policies become the optimal solution among any set of policies.

We note that the Stackelberg equilibria can be obtained from the analysis presented by Tamura \cite[Theorem 2]{Tamura2018} which characterizes linear policies that form a Stackelberg equilibrium for a general quadratic multidimensional signaling setup for jointly Gaussian sources with some generalizations (in the cost setup considered) but also some restrictions, such as the \textit{a priori} restriction of the decoder to an affine class in the conditional expectation (that limits the applicability for the noisy channel setup to be considered later in the paper.) Accordingly, we consider an alternative approach where we use the equivalent formulation employed in Section~\ref{sec:Nash}.

\begin{theorem}\label{thm:StackelbergVector}
\begin{enumerate}
\item[(i)] The encoding policies in \eqref{eq:NashEnc} with $\alpha_i\neq 0$ for all $i=1,\dots,\ySize$ and the decoding policies in \eqref{eq:NashDec} form a Stackelberg equilibrium, i.e., a payoff dominant Nash equilibrium and a Stackelberg equilibrium coincide.
\item[(ii)] In contrast to Nash setup for which there exist both informative and noninformative equilibria, a Stackelberg equilibrium is always informative, where the sender uses the private and nonprivate random variables in constructing its message.
\item[(iii)] When the nonprivate random variable is not scalar valued, i.e., $\ySize>1$, there exist informative Nash equilibria which do not coincide with the Stackelberg equilibria. These Nash equilibria are attained by an encoding policy in \eqref{eq:NashEnc} and a decoding policy in \eqref{eq:NashDec} where the scalars $\{\alpha_i\}_{i=1}^{\ySize}$ can take any value with at least one zero term and one nonzero term.
\end{enumerate}
\end{theorem}

\begin{IEEEproof}
We apply a transformation of variables as in \eqref{eq:defT}. In this transformed coordinate system, the objective functions of the sender and the receiver are expressed as in \eqref{eq:JeEqui} and \eqref{eq:JdEqui}, respectively. If we partition the random vector $\boldsymbol{T}$ in this coordinate system as in \eqref{eq:defUandV} according to the sign of coefficients in \eqref{eq:JeEqui}, we can show that the sender can only convey information related to $\boldsymbol{U}$. In particular, in Lemma~\ref{lem:DoNotSendVStackelberg} of Appendix~\ref{sec:supporting}, we prove that the sender does not convey information related to $\boldsymbol{V}$ at a Stackelberg equilibrium, which is proven in a similar manner to the proof of Lemma~\ref{lem:DoNotSendVNash} with the exception that in this case the sender announces its policy first.

As a result of Lemma~\ref{lem:DoNotSendVStackelberg}, the sender is restricted to transmit $\boldsymbol{U}$. Since the linear encoding policies in \eqref{eq:NashEnc} with $\alpha_i\neq 0$ for all $i=1,\dots,\ySize$ reveals $\boldsymbol{U}$ completely, these encoding policies yield the minimum attainable performance for the sender among all encoding policies. Therefore, the policies in the statement of the theorem lead to a Stackelberg equilibrium. 

We note that the random variable $\boldsymbol{U}$, which is desired to be conveyed at a Stackelberg equilibrium, has a size of $\ySize\geq 1$. Therefore, there always exists informative Stackelberg equilibrium where the sender reveals $\boldsymbol{U}$ completely. On the other hand, it is always possible to construct a noninformative Nash equilibrium as noted in Theorem~\ref{thm:NashVector}. 

When $\ySize>1$, the random variable $\boldsymbol{U}$ is not scalar valued. Thus, there exist informative Nash equilibria with an encoding policy in \eqref{eq:NashEnc} where the scalars $\{\alpha_i\}_{i=1}^{\ySize}$ take any value with at least one zero term and one nonzero term. In these informative Nash equilibria, the sender conveys only a subset of the random variables in $\boldsymbol{U}$. Since at these informative Nash equilibria, the performance of the sender is strictly worse than that at the payoff dominant Nash equilibria, these policies do not lead to a Stackelberg equilibrium. 
\end{IEEEproof}

\begin{remark}
As noted earlier, \cite{Farokhi2015QuadraticGames} considers a Stackelberg game setup where there is also side information at the receiver. In particular, \cite[Theorem~3]{Farokhi2015QuadraticGames} makes an \emph{a priori} affine policy restriction and provides an implicit characterization for the equilibrium solution in terms of an optimization problem. On the other hand, Theorem~\ref{thm:StackelbergVector} of our manuscript does not make an \emph{a priori} affine policy restriction and provides an explicit characterization of the Stackelberg equilibrium solution. In that sense, our result reveals that linear policies arise as the equilibrium solution among any set of policies for the Stackelberg game setup without receiver side information. In addition, if we consider the optimization problem that specifies equilibrium policies under the affine policy restriction in \cite[Theorem~3]{Farokhi2015QuadraticGames} and ignore the receiver side information, then the policies in Theorem~\ref{thm:StackelbergVector} of our manuscript give the optimal solution, as expected. 
\end{remark}

\begin{remark}
It is interesting to contrast our results in the case of a privacy concerned sender with the results in the strategic information transmission literature involving a biased sender. If one considers the classical setup of Crawford and Sobel \cite{CrawfordSobel} with a biased sender under the Stackelberg equilibrium concept (rather than the Nash equilibrium concept as investigated in \cite{CrawfordSobel}), then there exist linear equilibria \cite{Saritas2017QuadraticEquilibria}. In addition, \cite{Ekyol2017ProcIEEE} considers a Gaussian signaling game setup with a biased sender where the bias is modeled as a random variable and shows the optimality of linear policies for the scalar case. Hence, similar to our setup with a privacy concerned sender, in the setups with a biased sender, the Stackelberg equilibrium solutions are given by linear policies in the case of Gaussian sources. 
\end{remark}

\begin{remark} 
In fact, employing any invertible function $h(\boldsymbol{u})$ at the sender also yields a Stackelberg equilibrium. Since the receiver knows the commitment of the sender, it can simply employ $h^{-1}(\boldsymbol{u})$ to perfectly recover $\boldsymbol{u}$.
\end{remark}

\section{The MMSE Gaussian Information Bottleneck Problem}\label{sec:IB}

We now visit and re-formulate the information bottleneck problem \cite{tishby1999information} as depicted in Figure \ref{fig:blockDiagramIB}. This problem has gained a significant attention in the recent literature \cite{IBtoPrivacyFunnel2014,TishbyITW2015,VeraISIT2018,VeraIT2019,GoldfeldJSAIT2020,HsuISIT2018,ZaidiEntropy2020,HarremoesISIT2007,Gilad-Bachrach2003,Dhillon2003}. We will interpret this problem as an instance of our formulation under the Stackelberg equilibrium concept in the following sense: in contrast to the privacy game setup, only the random variable $\boldsymbol{X}$ is observed at the sender in the information bottleneck setup. In particular, we provide an estimation theoretic perspective on the information bottleneck problem by using quadratic distortion criteria as in our privacy game setup. 

The information bottleneck problem considers a similar objective to that employed in this manuscript where the performance metrics involve mutual information rather than the mean squared error. In the information bottleneck problem, the aim is to compress an observed random variable while trying to preserve information related to a correlated random variable. These conflicting objectives are analyzed by formulating an optimization problem involving the mutual information between the considered random variables. Although the problem can be cast as a constrained optimization problem of maximizing the released (useful) information related to the unobserved random variable under a compression constraint, the Lagrangian dual of this constrained problem is commonly considered (see Remark~\ref{rem:constrainedIB} for a constrained version in our setting). The aim is to find the optimal solution to the following optimization problem:
\begin{align}
\min_{\boldsymbol{Z}=\gamma^e(\boldsymbol{X})} I(\boldsymbol{X};\boldsymbol{Z})-\beta I(\boldsymbol{Y};\boldsymbol{Z}), \label{eq:IBObjective}
\end{align}
where $\beta\geq 0$ is a tradeoff parameter. Here, the goal is to compress an observation $\boldsymbol{X}$ while at the same time to maximize the released information related to $\boldsymbol{Y}$. 

The information bottleneck problem considering jointly Gaussian multidimensional sources is studied in \cite{chechik2005IB} where the structure of the optimal solution, which is jointly Gaussian with $\boldsymbol{X}$ \cite{IBOptimality}, is identified. The objective function in \eqref{eq:IBObjective} resembles the objective function considered in this manuscript in the sense that in both problems the random variable $\boldsymbol{Y}$ is desired to be conveyed while information related to the random variable $\boldsymbol{X}$ is desired to be removed from the displayed message. 

The information bottleneck problem can in fact be viewed as a Stackelberg game between a sender and a receiver. In this game, the sender wants to compress the observed random variable and to convey the unobserved random variable. The use of mutual information as a performance metric effectively means that the receiver uses all the available information related to both of the random variables, i.e., it always employs its best response as in the Stackelberg equilibrium. Thus, the receiver is concerned with extracting information related to both of the random variables, which is also the case in our framework. 

In the following, we consider a setting which is similar to the information bottleneck technique by employing mean squared error terms as our metric. As in the information bottleneck framework, the sender observes only the random variable $\boldsymbol{X}$, rather than observing both of the random variables. Namely, the encoder has access to only partial information and is of the form $\boldsymbol{z}=\gamma^e(\boldsymbol{x})$. The objective functions of the sender and receiver are as defined in \eqref{eq:objEnc} and \eqref{eq:objDec}, respectively. Since the receiver is concerned with estimating both of the random variables, it employs the minimum mean squared error estimators of each random variable. Since the equilibrium concept is the Stackelberg equilibrium, the objective function of the sender can be written as 
\begin{align}\label{eq:IBObjective2}
J^e(\gamma^e) = 
-\delta \, \Expect[\lVert \boldsymbol{X}-\Expect[\boldsymbol{X}|\boldsymbol{Z}]\rVert^2] + \Expect[\lVert \boldsymbol{Y}-\Expect[\boldsymbol{Y}|\boldsymbol{Z}]\rVert^2].
\end{align}
We now present the MMSE Gaussian information bottleneck solution.

\begin{theorem}\label{thm:IB}
\begin{enumerate}
\item[(i)] When $\Sigma_{\boldsymbol{X}\boldsymbol{Y}}\Sigma_{\boldsymbol{Y}\boldsymbol{X}}-\delta \Sigma_{\boldsymbol{X}}^2$ is not negative definite, the MMSE Gaussian information bottleneck solution is attained by an encoding policy
\begin{align}\label{eq:IBEnc}
\boldsymbol{z} = 
\begin{bmatrix}
\alpha_1\boldsymbol{q}_1 & \dots & \alpha_k \boldsymbol{q}_k
\end{bmatrix}^T
\Sigma_{\boldsymbol{X}}^{-1/2} \boldsymbol{x}
\end{align}
for nonzero scalars $\{\alpha_i\}_{i=1}^k$ where $k$ denotes the number of nonnegative eigenvalues of $W = \Sigma_{\boldsymbol{X}}^{-1/2}\Sigma_{\boldsymbol{X}\boldsymbol{Y}}\Sigma_{\boldsymbol{Y}\boldsymbol{X}}\Sigma_{\boldsymbol{X}}^{-1/2}-\delta \Sigma_{\boldsymbol{X}}$ and $\{\boldsymbol{q}_i\}_{i=1}^k$ are the normalized eigenvectors of $W$ associated with its nonnegative eigenvalues. The corresponding decoding policy is given by
\begin{align}\label{eq:IBDec}
\begin{bmatrix}
\gamma^{d_{\boldsymbol{X}}}(\boldsymbol{z})\\ \gamma^{d_{\boldsymbol{Y}}}(\boldsymbol{z})
\end{bmatrix}
= &\mathrm{diag}(\Sigma_{\boldsymbol{X}}^{1/2} ,\Sigma_{\boldsymbol{Y}\boldsymbol{X}}\Sigma_{\boldsymbol{X}}^{-1/2})
\nonumber \\
&\times 
\begin{bmatrix}
\beta_1\boldsymbol{q}_1 & \cdots & \beta_k\boldsymbol{q}_k 
\end{bmatrix}
\boldsymbol{z}
\end{align}
where $\beta_i = 1/\alpha_i$ for $i=1,\dots,k$.
\item[(ii)] In the particular case when $\Sigma_{\boldsymbol{X}\boldsymbol{Y}}\Sigma_{\boldsymbol{Y}\boldsymbol{X}}-\delta \Sigma_{\boldsymbol{X}}^2$ is positive definite, the MMSE Gaussian information bottleneck solution is attained by a fully informative encoding policy, where the sender reveals the random variable $\boldsymbol{X}$ completely.
\item[(iii)] When $\Sigma_{\boldsymbol{X}\boldsymbol{Y}}\Sigma_{\boldsymbol{Y}\boldsymbol{X}}-\delta \Sigma_{\boldsymbol{X}}^2$ is negative definite, the MMSE Gaussian information bottleneck solution is noninformative, i.e., the sender does not reveal any information related to its observation. 
\end{enumerate}
\end{theorem}

Before we present the proof, we contrast our estimation theoretic solution of information bottleneck problem in Theorem~\ref{thm:IB} with the information theoretic solution in \cite{chechik2005IB}. Towards that goal, we restate \cite[Theorem~3.1]{chechik2005IB} which gives the solution to the optimization problem in (\ref{eq:IBObjective}).

\begin{theorem}[{\cite[Theorem~3.1]{chechik2005IB}}]
For the Gaussian information bottleneck problem under (\ref{eq:IBObjective}), the optimal solution is given by 
\begin{align}\label{eq:IBSolutionIT}
\boldsymbol{z} = 
A(\beta)
\boldsymbol{x}
+ 
\boldsymbol{n}
\end{align}
where $\boldsymbol{n}$ is a realization of a zero-mean Gaussian random vector with identity covariance matrix and
\begin{align*}
A(\beta) = 
\begin{cases}
[
\boldsymbol{0} \;\, \cdots \;\, \boldsymbol{0}
]^T
& \text{if }0\leq \beta< \beta_1^c\\
[
\alpha_1 \boldsymbol{p}_1 \;\, \boldsymbol{0} \;\,\cdots \;\, \boldsymbol{0}
]^T
& \text{if }\beta_1^c \leq \beta< \beta_2^c\\
[
\alpha_1 \boldsymbol{p}_1 \;\, \alpha_2 \boldsymbol{p}_2 \;\, \boldsymbol{0} \;\,\cdots \;\boldsymbol{0}
]^T
& \text{if }\beta_2^c \leq \beta< \beta_3^c\\
\quad \vdots
\end{cases}
\end{align*}
where $\{\boldsymbol{p}_i^T\}_{i=1}^{\xSize}$ are left eigenvectors of 
\[\Sigma_{\boldsymbol{X}|\boldsymbol{Y}}\Sigma_{\boldsymbol{X}}^{-1}\triangleq (\Sigma_{\boldsymbol{X}}-\Sigma_{\boldsymbol{X}\boldsymbol{Y}}\Sigma_{\boldsymbol{Y}}^{-1}\Sigma_{\boldsymbol{Y}\boldsymbol{X}})\Sigma_{\boldsymbol{X}}^{-1}\] with the corresponding eigenvalues $\{\lambda_i\}_{i=1}^{\xSize}$, which are sorted in ascending order, $\beta_i^c=1/(1-\lambda_i)$ and $\alpha_i = \sqrt{\frac{\beta(1-\lambda_i)-1}{\lambda_i (\boldsymbol{p}_i^T\Sigma_{\boldsymbol{X}}\boldsymbol{p}_i)}}$ for $i=1,\dots,\xSize$.
\end{theorem}

\begin{remark}
We note that for the information bottleneck problem under (\ref{eq:IBObjective}), the optimal solution is jointly Gaussian with $\boldsymbol{X}$ which is also the case for our estimation theoretic solution in Theorem~\ref{thm:IB}. However, in the information theoretic formulation, the solution involves applying an independent Gaussian perturbation to a linear function of $\boldsymbol{X}$ as given in \eqref{eq:IBSolutionIT} whereas in our estimation theoretic solution the encoder reveals a linear function of $\boldsymbol{X}$ without applying any perturbation. We note that if one considers the original unconstrained version of the information bottleneck problem, then randomization will be needed to ensure that the constraint condition holds with an equality in some constraint regime. See Remark~\ref{rem:constrainedIB} for further details.
\end{remark}

Next, we present the proof of Theorem~\ref{thm:IB}.

\begin{IEEEproof} We have that $(\boldsymbol{Y}-\Expect[\boldsymbol{Y}|\boldsymbol{X}])$ is orthogonal to $\boldsymbol{X}$ since \[\Expect[\boldsymbol{Y}\boldsymbol{X}] = \Expect[\Expect[\boldsymbol{Y}\boldsymbol{X} |\boldsymbol{X}]] = \Expect[ \Expect[\boldsymbol{Y}|\boldsymbol{X}]\boldsymbol{X}].\] Since the sender observes only the random variable $\boldsymbol{X}$ and determines its message based on $\boldsymbol{X}$, it follows that $\boldsymbol{Y}\rightarrow \boldsymbol{X}\rightarrow \boldsymbol{Z}$ is a Markov chain in that order. This Markov property implies that $(\boldsymbol{Y}-\Expect[\boldsymbol{Y}|\boldsymbol{X}])$ is orthogonal to $\boldsymbol{Z}$ as well. To see this, observe that 
\[\Expect[\boldsymbol{Z}\Expect[\boldsymbol{Y}|\boldsymbol{X}] ] = \Expect[\boldsymbol{Z}  \Expect[\boldsymbol{Y}|\boldsymbol{X},\boldsymbol{Z}] ] =  \Expect[\Expect[\boldsymbol{Z}\boldsymbol{Y}|\boldsymbol{X}, \boldsymbol{Z}] ] = \Expect[\boldsymbol{Z}\boldsymbol{Y}]\]
 where the first equality is due to the Markov property and the last equality is due to iterated expectations. By using these orthogonality properties, we can express the second term in \eqref{eq:IBObjective2} as
\begin{align*}
&\Expect[\lVert\boldsymbol{Y}-\Expect[\boldsymbol{Y}|\boldsymbol{Z}]\rVert^2] \nonumber \\
&= \Expect[\lVert \boldsymbol{Y}-\Expect[\boldsymbol{Y}|\boldsymbol{X}]+\Expect[\boldsymbol{Y}|\boldsymbol{X}]-\Expect[\boldsymbol{Y}|\boldsymbol{Z}]\rVert^2] \nonumber \\
&= \Expect[\lVert \boldsymbol{Y}-\Expect[\boldsymbol{Y}|\boldsymbol{X}]\rVert^2]+\Expect[\lVert\Expect[\boldsymbol{Y}|\boldsymbol{X}]-\Expect[\boldsymbol{Y}|
\boldsymbol{Z}]\rVert^2] \nonumber\\
&= \Expect[\lVert \boldsymbol{Y}-\Expect[\boldsymbol{Y}|\boldsymbol{X}]\rVert^2]
+\Expect[\lVert\Expect[\boldsymbol{Y}|\boldsymbol{X}]-\Expect[\Expect[\boldsymbol{Y}|\boldsymbol{X}]|\boldsymbol{Z}]\rVert^2] 
\end{align*}
where the second equality follows from the fact that $(\boldsymbol{Y}-\Expect[\boldsymbol{Y}|\boldsymbol{X}])$ is orthogonal to $\boldsymbol{X}$ and $\boldsymbol{Z}$ and the third equality is due to iterated expectations. Observing that 
\[\Expect[\boldsymbol{Y}|\boldsymbol{X}=\boldsymbol{x}] = \Sigma_{\boldsymbol{Y}\boldsymbol{X}}\Sigma_{\boldsymbol{X}}^{-1}\boldsymbol{x},\] the objective function can be written as 
\begin{align*}
&J^e(\gamma^e) = 
\Expect[\lVert \boldsymbol{Y}-\Expect[\boldsymbol{Y}|\boldsymbol{X}]\rVert^2] 
-\delta \, \Expect[\lVert \boldsymbol{X}-\Expect[\boldsymbol{X}|\boldsymbol{Z}]\rVert^2] \nonumber \\
&+\Expect[
(\boldsymbol{X}-\Expect[\boldsymbol{X}|\boldsymbol{Z}])^T
(\Sigma_{\boldsymbol{X}}^{-1}\Sigma_{\boldsymbol{X}\boldsymbol{Y}}\Sigma_{\boldsymbol{Y}\boldsymbol{X}}\Sigma_{\boldsymbol{X}}^{-1})
(\boldsymbol{X}-\Expect[\boldsymbol{X}|\boldsymbol{Z}])
],
\end{align*}
where the first term is independent of the encoder. Thus, we obtain an optimization problem of the form
\begin{align}\label{eq:IBReduced}
\min_{\boldsymbol{Z}=\gamma^e(\boldsymbol{X})}
\Expect\Big[ 
(\boldsymbol{X}-\Expect[\boldsymbol{X}|\boldsymbol{Z}])^T
M
(\boldsymbol{X}-\Expect[\boldsymbol{X}|\boldsymbol{Z}])
\Big],
\end{align}
where \[M\triangleq (\Sigma_{\boldsymbol{X}}^{-1}\Sigma_{\boldsymbol{X}\boldsymbol{Y}}\Sigma_{\boldsymbol{Y}\boldsymbol{X}}\Sigma_{\boldsymbol{X}}^{-1}-\delta I).\] The optimization problem in \eqref{eq:IBReduced} can be viewed as a quadratic multidimensional signaling game problem considered earlier in the paper and the solution can be obtained by using the analysis in Section \ref{sec:Stackelberg}. In particular, we can rewrite the problem in \eqref{eq:IBReduced} as a Stackelberg game between a sender and a receiver with objective functions
\begin{align}\label{eq:EncObjIB}
J^e(\gamma^e,\gamma^{d_{\boldsymbol{X}}}) 
&= \Expect[
(\boldsymbol{X}-\gamma^{d_{\boldsymbol{X}}}(\boldsymbol{Z}))^T
M
(\boldsymbol{X}-\gamma^{d_{\boldsymbol{X}}}(\boldsymbol{Z}))
],
\\
\label{eq:DecObjIB}
J^d(\gamma^e,\gamma^{d_{\boldsymbol{X}}}) 
&= \Expect[
(\boldsymbol{X}-\gamma^{d_{\boldsymbol{X}}}(\boldsymbol{Z}))^T
(\boldsymbol{X}-\gamma^{d_{\boldsymbol{X}}}(\boldsymbol{Z}))
].
\end{align}
Notice that for a given encoding policy, the best response of the receiver under \eqref{eq:DecObjIB} is given by the minimum mean squared estimator of $\boldsymbol{X}$ given $\boldsymbol{Z}$, which is consistent with \eqref{eq:IBReduced}.

Next, we apply a transformation of variables to express the objective function of the sender in a simplified form. Towards that goal, let $W = \Sigma_{\boldsymbol{X}}^{-1/2}\Sigma_{\boldsymbol{X}\boldsymbol{Y}}\Sigma_{\boldsymbol{Y}\boldsymbol{X}}\Sigma_{\boldsymbol{X}}^{-1/2}-\delta \Sigma_{\boldsymbol{X}} = Q \Lambda Q^T$ for orthonormal $Q$ and diagonal $\Lambda$. Now, consider the following invertible transformation of variables
\begin{align}\label{eq:transformationIB}
\boldsymbol{T} \triangleq Q^T \Sigma_{\boldsymbol{X}}^{-1/2} \boldsymbol{X}.
\end{align}
Under this transformation of variables, we introduce an equivalent problem in a similar manner to Section~\ref{sec:Nash}. The observation $\boldsymbol{X}$ is mapped into $\boldsymbol{T}$ via a fixed transformation \eqref{eq:transformationIB}. The encoder chooses an arbitrary policy $\tilde{\gamma}^e(\cdot)$ which maps the transformed random vector $\boldsymbol{T}$ into the message $\boldsymbol{Z}$. The receiver applies an arbitrary policy $\gamma^{d_{\boldsymbol{T}}}(\cdot)$ to its observation $\boldsymbol{Z}$ and then the estimate of $\boldsymbol{X}$ is obtained via a fixed relation 
\begin{align}
\gamma^{d_{\boldsymbol{X}}}(\boldsymbol{z}) = \Sigma_{\boldsymbol{X}}^{1/2} Q \gamma^{d_{\boldsymbol{T}}}(\boldsymbol{z}).
\end{align}

In this transformed coordinate system, for a given encoding policy the receiver still employs the minimum mean squared estimator of the random variable $\boldsymbol{T}$, which can be established via a similar analysis to that in Lemma~\ref{lem:MMSEOptimal}. The objective function of the sender in this transformed coordinate system can be written
\begin{align}\label{eq:IBEncReduced}
J^e(\tilde{\gamma}^e,\gamma^{d_{\boldsymbol{T}}})
= \sum_{i=1}^{\xSize}
\lambda_i 
\Expect[
(T_i-\gamma^{d_{T_i}}(\boldsymbol{Z}))^2
],
\end{align} 
where $\{\lambda_i\}_{i=1}^{\xSize}$ are eigenvalues of $W$. 

If all of these eigenvalues $\{\lambda_i\}_{i=1}^{\xSize}$ are positive, which is equivalent to $\Sigma_{\boldsymbol{X}\boldsymbol{Y}}\Sigma_{\boldsymbol{Y}\boldsymbol{X}}-\delta \Sigma_{\boldsymbol{X}}^2$ being positive definite, then the minimum can be attained by revealing $\boldsymbol{T}$, which corresponds to the fully informative scenario. In case all of these eigenvalues are negative, i.e., $\Sigma_{\boldsymbol{X}\boldsymbol{Y}}\Sigma_{\boldsymbol{Y}\boldsymbol{X}}-\delta \Sigma_{\boldsymbol{X}}^2$ is negative definite, then revealing information related to any component of $\boldsymbol{T}$ is not desirable for the sender, and thus, this scenario leads to a noninformative Stackelberg equilibrium. In the remaining case, i.e., $\Sigma_{\boldsymbol{X}\boldsymbol{Y}}\Sigma_{\boldsymbol{Y}\boldsymbol{X}}-\delta \Sigma_{\boldsymbol{X}}^2$ is neither positive definite nor negative definite, we partition the transformed vector according to the sign of the coefficients $\{\lambda_i\}_{i=1}^{\xSize}$ in \eqref{eq:IBEncReduced} as follows:
\begin{align}
\boldsymbol{T} \triangleq \begin{bmatrix}
\boldsymbol{U}\\
\boldsymbol{V}
\end{bmatrix}
\end{align}
where $\boldsymbol{U}\in\mathbb{R}^k$ and $\boldsymbol{V}\in\mathbb{R}^{\xSize-k}$ correspond to nonnegative and negative coefficients in \eqref{eq:IBEncReduced}, respectively, with $k$ denoting the number of nonnegative eigenvalues of $W$. Next, we can apply Lemma~\ref{lem:DoNotSendVStackelberg} for this particular Stackelberg game setup to establish that the sender cannot convey information related to $\boldsymbol{V}$ and is restricted to send information related to $\boldsymbol{U}$. As the encoding policy in \eqref{eq:IBEnc} reveals $\boldsymbol{U}$ completely, this encoding policy achieves the minimum attainable for the sender among any set of policies. Thus, the pair of policies \eqref{eq:IBEnc} and \eqref{eq:IBDec} yield a Stackelberg equilibrium, which gives the solution to the MMSE Gaussian information bottleneck problem. 
\end{IEEEproof}

\begin{remark}
As indicated in Theorem~\ref{thm:IB}, in the information bottleneck setup, the solution may be informative or noninformative depending on the tradeoff parameter $\delta$. In contrast, in the privacy-signaling setup investigated in Section~\ref{sec:Stackelberg}, the equilibrium solution is always informative regardless of $\delta$ as shown in Theorem~\ref{thm:StackelbergVector}. We can have the following interpretation regarding these results: In the privacy-signaling setup, the sender can perform perfect removal of information in the revealed message according to its objective as it has access to both of the random variables. It turns out that this is attained via a linear encoding policy for the case of Gaussian sources. On the other hand, in the information bottleneck setup, the sender having access to partial information cannot apply perfect information removal. Instead, the sender does what is best given the partial information that it has. This happens to be a full disclosure, a partial disclosure, or a no disclosure policy depending on $\delta$. 
\end{remark}

In the special case of scalar sources, Theorem~\ref{thm:IB} simplifies. In particular, depending on the value of $\delta$, the equilibrium is either fully informative or noninformative and we summarize this result in the following corollary.

\begin{corollary}\label{thm:IBScalar}
The Stackelberg equilibrium of the information bottleneck problem for scalar sources is given by one of the following cases: 
\begin{enumerate}
\item[(i)] If $(\rho^2/\sigma_X^4)>\delta$, then the sender completely reveals $X$.
\item[(ii)] If $(\rho^2/\sigma_X^4)<\delta$, then the sender does not reveal information related to $X$.
\item[(iii)] If $(\rho^2/\sigma_X^4)=\delta$, then both informative and noninformative scenarios lead to a Stackelberg equilibrium.
\end{enumerate}
\end{corollary}

\begin{remark}\label{rem:constrainedIB}[{\bf Constrained MMSE Information Bottleneck Problem.}]
It is also possible to apply the ideas used in the proof of Theorem~\ref{thm:IB} to a constrained version of the MMSE Gaussian information bottleneck problem where the aim is to minimize the mean squared error for estimating $\boldsymbol{Y}$ under constraint that the mean squared error for estimating $\boldsymbol{X}$ is above a certain threshold $\alpha$. In this case, the problem is defined with
\begin{align}
&\min_{\boldsymbol{Z} = \gamma^e(\boldsymbol{X})} 
\mathrm{Tr}
\Big(\Upsilon\,\Expect[(\boldsymbol{X} - \Expect[\boldsymbol{X}|\boldsymbol{Z}]) 
(\boldsymbol{X} - \Expect[\boldsymbol{X}|\boldsymbol{Z}])^T 
]
\Big)\nonumber \\
&\subjectto 
\mathrm{Tr}
\Big( \Expect[(\boldsymbol{X} - \Expect[\boldsymbol{X}|\boldsymbol{Z}]) 
(\boldsymbol{X} - \Expect[\boldsymbol{X}|\boldsymbol{Z}])^T
]
\Big) \geq \alpha 
\end{align}
where $\Upsilon\triangleq \Sigma_{\boldsymbol{X}}^{-1}\Sigma_{\boldsymbol{X}\boldsymbol{Y}}\Sigma_{\boldsymbol{Y}\boldsymbol{X}}\Sigma_{\boldsymbol{X}}^{-1}$, which is always positive semidefinite. Since any positive semidefinite 
\[\Phi \triangleq \Expect[(\boldsymbol{X} - \Expect[\boldsymbol{X}|\boldsymbol{Z}]) 
(\boldsymbol{X} - \Expect[\boldsymbol{X}|\boldsymbol{Z}])^T ] \] is attainable via a linear encoding policy with a Gaussian perturbation, the problem reduces to 
\begin{align}
&\min_{\Phi\succeq 0} \;\mathrm{Tr}(\Upsilon \,\Phi) \nonumber \\
&\subjectto \mathrm{Tr}(\Phi)\geq \alpha.
\end{align}
Let the minimum eigenvalue of $\Upsilon$ be denoted by $\lambda_{\text{min}}$. Observe that
\begin{align*}
\mathrm{Tr}(\Upsilon\Phi) 
&= \mathrm{Tr}(\Upsilon\Phi-\lambda_{\text{min}}\Phi+\lambda_{\text{min}}\Phi)\\
&= \mathrm{Tr}((\Upsilon-\lambda_{\text{min}}I)\Phi) + \lambda_{\text{min}}\mathrm{Tr}(\Phi)\\
&\geq \alpha\lambda_{\text{min}}
\end{align*}
where the inequality uses the constraint along with the observation that $(\Upsilon-\lambda_{\text{min}}I)$ and $\Phi$ are positive semidefinite. As a result, by using a linear encoding policy possibly with a Gaussian perturbation, one can attain $\mathrm{Tr}(\Phi)=\alpha$ where the solution satisfies the orthogonality condition under the trace inner-product defining a Hilbert space on square matrices: 
\[\mathrm{Tr}((\Upsilon-\lambda_{\text{min}}I)\Phi) = 0.\] Since such an encoding policy achieves the characterized lower bound, it becomes the optimal solution to the constrained MMSE Gaussian information bottleneck problem. That the constrained problem with inequality is equivalent to a problem with an equality constraint applies more broadly to information bottleneck problems, see e.g. \cite{WitsenhausenIT1975}.
\end{remark}

\begin{remark}
We emphasize that the solution presented in Theorem~\ref{thm:IB} is obtained without making an \textit{a priori} linear policy restriction. These policies are the optimal solution among any set of policies for the optimization problem constructed at the sender by anticipating the best response of the receiver. 
\end{remark}

\begin{remark}
In the information bottleneck problem, the sender uses partial information since only random variable $\boldsymbol{X}$ is available at the sender whereas in our privacy-signaling game formulation the sender has access to both of the random variables. Due to this further restriction that only partial information is available at the sender, our information bottleneck analysis provides a lower bound on the performance of our original Stackelberg game setting.
\end{remark}

\begin{remark}
It should be emphasized that the information bottleneck problem involving mutual information corresponds to the Stackelberg equilibrium concept since employing mutual information effectively means that the receiver uses all the available information, i.e., it employs its best response. On the other hand, the Nash problem would require an explicit dependence of the functions (considered in the optimization) on the receiver policy. 
\end{remark}

\section{A Channel between the Sender and the Receiver} \label{sec:channel}

In this section, we generalize our results on the considered privacy-signaling game problem to scenarios when there is a channel between the sender and the receiver. In fact, the proposed equivalent formulation employed in the proof of Theorem~\ref{thm:NashVector} is also applicable when there is a channel between the sender and the receiver. Namely, Lemma~\ref{lem:DoNotSendVNash} and Lemma~\ref{lem:DoNotSendVStackelberg} generalize to scenarios with a channel between the players represented by a conditional distribution $p(\boldsymbol{r}|\boldsymbol{z})$ where $\boldsymbol{R}=\boldsymbol{r}$ denotes the observation of the receiver. These generalizations imply that the sender cannot transmit information related to $\boldsymbol{V}$ and is restricted to send information related to $\boldsymbol{U}$ at a Nash equilibrium or at a Stackelberg equilibrium, where $\boldsymbol{U}$ and $\boldsymbol{V}$ are partitions of the transformed coordinate system specified in \eqref{eq:defUandV}. Thus, for a given channel, the aim is to find an encoder/decoder pair that is optimal in conveying a sequence of independent zero-mean Gaussian distributed sources over that particular channel in a mean squared error sense and such an optimal encoding/decoding policy pair leads to a payoff dominant Nash equilibrium as well as a Stackelberg equilibrium. 

In the following, we focus on the particular case of scalar sources and investigate the Nash and the Stackelberg equilibria for two important channel settings.

\subsection{Gaussian Noise Channel between the Sender and the Receiver}\label{sec:noisy}

In this subsection, we consider the same problem as before for scalar sources except that there is an additive Gaussian noise (e.g., measurement noise) between the transmitter and the receiver. More specifically, the sender encodes $X$ and $Y$ into $Z$ which is subject to additive noise $W$ and the receiver uses the observation $R=Z+W$ while decoding both of the random variables. The additive noise term is independent of $X$ and $Y$ and it is modeled as zero-mean Gaussian with variance $\sigma_W^2$. In addition, we assume that there is an average power constraint at the sender, i.e., $\Expect[Z^2]\leq P$.

\subsubsection{Nash Equilibria}

\begin{theorem}\label{thm:NashNoisy}
\begin{enumerate}
\item[(i)] There exist informative linear Nash equilibria with an encoding policy $\gamma^e(x,y) = Ax+By$ that satisfies 
\begin{align}\label{eq:BARatioNoisy}
\frac{B}{A}= -
\frac{(\delta\sigma_X^2+\sigma_Y^2)+\sqrt{(\delta\sigma_X^2+\sigma_Y^2)^2-4\delta\rho^2}}
{2\delta \rho},
\end{align}
\begin{align}\label{eq:powerNoisy}
A^2\sigma_X^2 + B^2\sigma_Y^2 +2AB\rho = P
\end{align}
and with decoding policies
\begin{align}
&\gamma^{d_X}(r) =
\left(
\frac
{A\sigma_X^2+B\rho}
{P +\sigma_W^2 }
\right)\label{eq:DecXNashNoisy}
r,\\
&\gamma^{d_Y}(r) =
\left(
\frac
{A\rho + B\sigma_Y^2}
{P +\sigma_W^2 }
\right)
r. \label{eq:DecYNashNoisy}
\end{align} 
\item[(ii)] These informative equilibria are payoff dominant Nash equilibria and they are the only possible payoff dominant Nash equilibria. Moreover, these equilibria are unique among the affine class of policies.
\end{enumerate}
\end{theorem}

\begin{IEEEproof}
Lemma~\ref{lem:DoNotSendVNash} implies that the sender is restricted to convey $U$ which corresponds to information conveyed by the sender at the equilibria specified in Corollary~\ref{cor:NashScalar}. Then, it is easy to verify that sending $U$ after scaling up to the maximum available power level yields a Nash equilibrium. The decoding policies at this equilibrium are given by the minimum mean squared error estimators corresponding to each random variable. Since the observation $R$ is jointly Gaussian with $X$ and $Y$, the conditional expectation formula for Gaussian distributions can be employed to obtain \eqref{eq:DecXNashNoisy} and \eqref{eq:DecYNashNoisy} \cite[p. 155]{PoorBook}.

The proof for the payoff dominance property of the equilibria uses the observation that the performance of both players is determined by $\Expect[(U-\gamma^{d_U}(R))^2]$ at a Nash equilibrium. Since the source is scalar and the Gaussian noise is additive, we can employ the well-known result that the problem of transmitting a scalar Gaussian source over a scalar Gaussian channel under an average power constraint admits a unique solution with linear encoding scaled to satisfy the power constraint with equality (see e.g. \cite[p. 376]{YukselBook2013}). Hence, the result immediately follows.

As it can be shown that an encoding policy $\tilde{\gamma}^e(u) = Au + C$ with $\Expect[(AU)^2]< P$ cannot be a Nash equilibrium, it follows that the only possible informative affine Nash equilibria are attained by an encoding policy that satisfies \eqref{eq:BARatioNoisy} and \eqref{eq:powerNoisy}.
\end{IEEEproof}

\subsubsection{Stackelberg Equilibria}

\begin{theorem}\label{thm:StackelbergNoisy}
The Stackelberg equilibria coincide with the payoff dominant Nash equilibria characterized in Theorem~\ref{thm:NashNoisy}. These equilibria are unique among any set of policies. 
\end{theorem}

\begin{IEEEproof}
Lemma~\ref{lem:DoNotSendVStackelberg} implies that the encoder cannot convey $V$ and it can only use $U$ in constructing its message. As the objectives of each player then becomes the minimization of the mean squared error for estimating $U$, the optimal strategy of the sender is to employ an encoding policy which is linear in $U$ with an average power equal to $P$. Moreover, this encoding strategy is unique due to the fact that it is the unique solution to the problem of transmitting a scalar Gaussian source over a scalar Gaussian channel under an average power constraint \cite{YukselBook2013}.
\end{IEEEproof}

It is important to emphasize that the encoder is not restricted to be affine. Since the problem reduces to transmitting a scalar Gaussian source over a scalar Gaussian channel under an average power constraint, we obtain these linear policies as the optimal unique solution to this reduced problem.

\subsection{Discrete Noiseless Channel between the Sender and the Receiver}\label{sec:discreteChannel}

In this subsection, we consider scalar sources and investigate the discrete channel setting where the sender is restricted to transmit a discrete value, i.e., $Z\in\{0,\dots,M-1\}$ for some $M\geq 2$. We assume that the channel is noiseless, i.e., $R=Z$.

While investigating the discrete channel setting, we again employ the equivalent formulation which facilitates the analysis. Lemma~\ref{lem:DoNotSendVNash} and Lemma~\ref{lem:DoNotSendVStackelberg} imply that both players share the common objective of minimizing $\Expect[(U-\gamma^{d_U}(R))^2]$ under both of the equilibrium concepts. Since the sender is restricted to transmit discrete values, it is required to quantize $U$ at the sender. Since this would correspond to classical quantization, the existence of an optimal quantizer follows from the classical results in the literature, e.g., \cite{GrayNeuhoffQuantization}. Namely, there exist quantization bins and reconstruction points which minimize the corresponding mean squared error. Thus, by assigning each bin to a discrete value of $Z$ and then using the corresponding optimal reconstruction points at the receiver yield a Nash equilibrium. We summarize this result in the following theorem.

\begin{theorem}\label{thm:NashDiscrete}
Consider the quantization of $U$ into $M$ bins where each bin is assigned to a discrete value of $Z$ at the encoder and the corresponding reconstruction points at the receiver such that $\Expect[(U-\gamma^{d_U}(R))^2]$ is minimized. This pair of encoding and decoding policies, which always exists, forms an informative Nash equilibrium. In addition, this equilibrium is a payoff dominant Nash equilibrium.
\end{theorem}

It is worth pointing out that for any number of bins lower than $M$, there exists a Nash equilibrium. In other words, even if $Z$ can take $M$ discrete values, a quantization policy using lower than $M$ bins at the sender and the corresponding reconstruction points at the receiver is also a Nash equilibrium. In addition, the case of a single bin is also a Nash equilibrium where no information related to $U$ is conveyed to the receiver. 

It is noted that using a large number of bins yields a lower $\Expect[(U-\gamma^{d_U}(R))^2]$. Since the mean squared error for estimating $U$ is desired to be minimized for both players, using a large number of bins results in improved objectives for both players. This monotonicity property with respect to the number of bins implies that at the Stackelberg equilibrium there must be $M$ bins. 

\begin{theorem}\label{thm:StackelbergDiscrete}
The pair of policies in Theorem~\ref{thm:NashDiscrete} leads to a Stackelberg equilibrium. 
\end{theorem}

\section{Numerical Examples}\label{sec:nume}

In this section, we provide numerical examples for the proposed privacy-signaling game and the MMSE information bottleneck problems.

\subsection{Scalar Sources} 
Here, we consider scalar sources and illustrate the performances at the characterized equilibria where the variances of the private and nonprivate random variables are set as $\sigma_X^2=\sigma_Y^2=1$. We consider only the privacy-signaling game problem since the information bottleneck solution is simply given by the fully informative or noninformative solution depending on $\delta$ in the case of scalar sources as stated in Corollary~\ref{thm:IBScalar}. Since the informative Nash equilibrium coincides with the Stackelberg equilibrium in the case of scalar sources for the privacy-signaling game setup, we do not make a distinction between them.

\begin{figure}
\centering
\includegraphics[width=\linewidth]{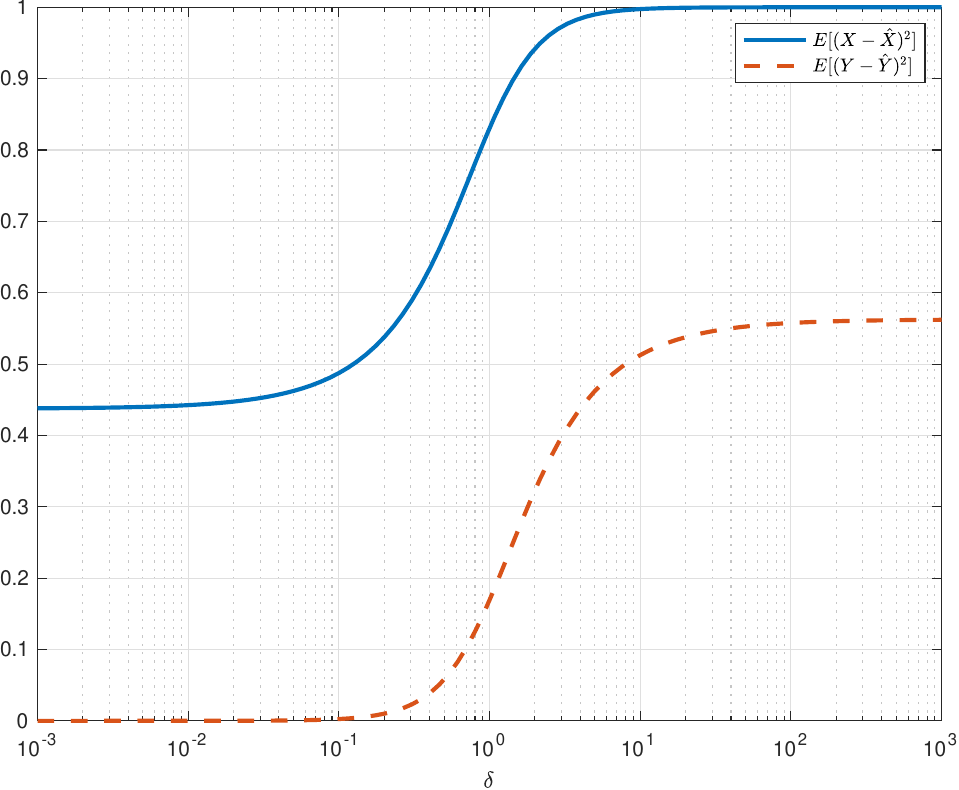}
\caption{Mean squared errors at the informative equilibrium for privacy-signaling game setup with respect to privacy ratio where $\rho=0.75$.}
\label{fig:scalar_delta_vs_cost}
\end{figure}

Fig.~\ref{fig:scalar_delta_vs_cost} plots the estimation errors for the private and nonprivate random variables with respect to the privacy ratio where the correlation between them is given by $\rho=0.75$. The estimation error for the private random variable increases with the privacy ratio since the transmitter removes information related to the private random variable due to enhanced privacy concerns. This removal also distorts the information conveyed related to the nonprivate random variable and hence the corresponding estimation error also increases.

\begin{figure}
\centering
\null\hfill
\begin{subfigure}[t]{0.48\textwidth}
\centering
\includegraphics[width=\textwidth]{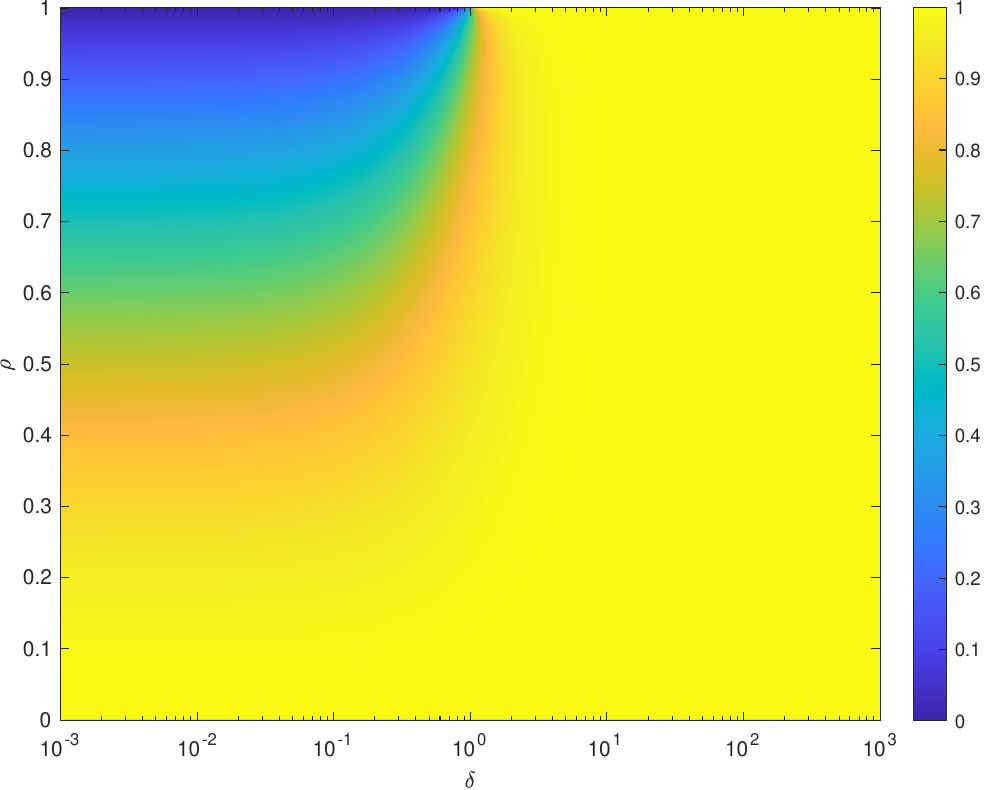}
\caption{Private random variable.}
\label{fig:scalar_imagesc_private}
\end{subfigure}
\hfill\vspace{5pt}
\begin{subfigure}[t]{0.48\textwidth}
\centering
\includegraphics[width=\textwidth]{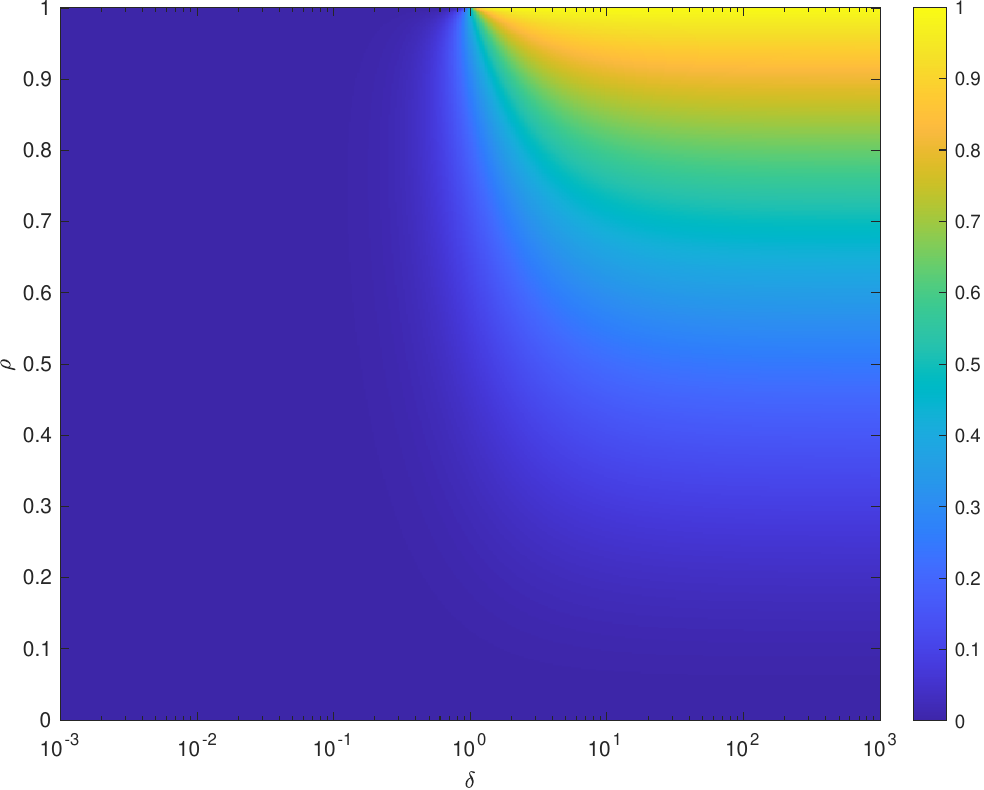}
\caption{Nonprivate random variable.}
\label{fig:scalar_imagesc_nonprivate}
\end{subfigure}
\hfill\null
\caption{Mean squared errors at the informative equilibrium for privacy-signaling game setup with respect to privacy ratio and correlation between the random variables.}
\label{fig:scalar_imagesc}
\end{figure}

Next, we illustrate the attained costs with respect to both the privacy ratio $\delta$ and the correlation between the random variables $\rho$. We plot the estimation errors at the equilibria in Fig.~\ref{fig:scalar_imagesc_private} for the private random variable and in Fig.~\ref{fig:scalar_imagesc_nonprivate} for the nonprivate random variable. In the low privacy scenario, the estimation error for $Y$ does not change significantly with respect to the correlation since most of the information contained in $Y$ is conveyed to the receiver regardless of the correlation. As a result, more information is leaked related to the private random variable as the correlation is increased. In contrast, in the high privacy scenario, regardless of the correlation, most of the information related to $X$ is removed from the transmitted message. Thus, the estimation error for the nonprivate random variable increases with the correlation whereas no significant changes in the estimation error for the private random variable are observed.

\begin{table}
\caption{The ratio of coefficients at the encoder for the derived informative equilibria when $\sigma_X^2=1$ and $\sigma_Y^2=1$, as specified in \eqref{eq:BARatio}. }
\label{tab:ratOfCoeff}
\begin{center}
\begin{tabular}{ l|c } 
Scenario & $B/A$ \\ \hline  
$\rho = 0.3$ and $\delta = 0.1$ & $-36.39$ \\\hline  
$\rho = 0.3$ and $\delta = 1$ & $-6.51$ \\\hline  
$\rho = 0.3$ and $\delta = 10$ & $-3.63$ \\\hline  
$\rho = 0.7$ and $\delta = 0.1$  & $-15.04$ \\\hline  
$\rho = 0.7$ and $\delta = 1$  & $-2.44$ \\\hline  
$\rho = 0.7$ and $\delta = 10$  & $-1.50$ \\
\end{tabular}
\end{center}
\end{table}

Table~\ref{tab:ratOfCoeff} illustrates the tradeoff between \textit{utility} in terms of conveying $Y$ and \textit{privacy} in terms of hiding $X$ by providing the structure of the encoder at the equilibrium for various values of the privacy ratio and correlation. It can be inferred that if the privacy ratio is increased while the correlation is kept the same, the information leakage related to the private random variable reduces, as expected.

\subsection{Multidimensional Sources}
Here, we consider vector valued sources where both the private and the nonprivate random variables are two-dimensional with the following covariance matrix:
\begin{align}
\Sigma = 
\begin{bmatrix}
\Sigma_{\boldsymbol{X}} & \Sigma_{\boldsymbol{X}\boldsymbol{Y}}\\
\Sigma_{\boldsymbol{Y}\boldsymbol{X}} & \Sigma_{\boldsymbol{Y}}
\end{bmatrix}
=
\begin{bmatrix}
1 & 0.7 & 0.7 & 0.6 \\
0.7 & 1 & 0.2 & 0.5 \\
0.7 & 0.2 & 1 & 0.6 \\
0.6 & 0.5 & 0.6 & 1
\end{bmatrix}.
\end{align}

\begin{figure}
\centering
\null\hfill
\begin{subfigure}[t]{0.48\textwidth}
\centering
\includegraphics[width=\textwidth]{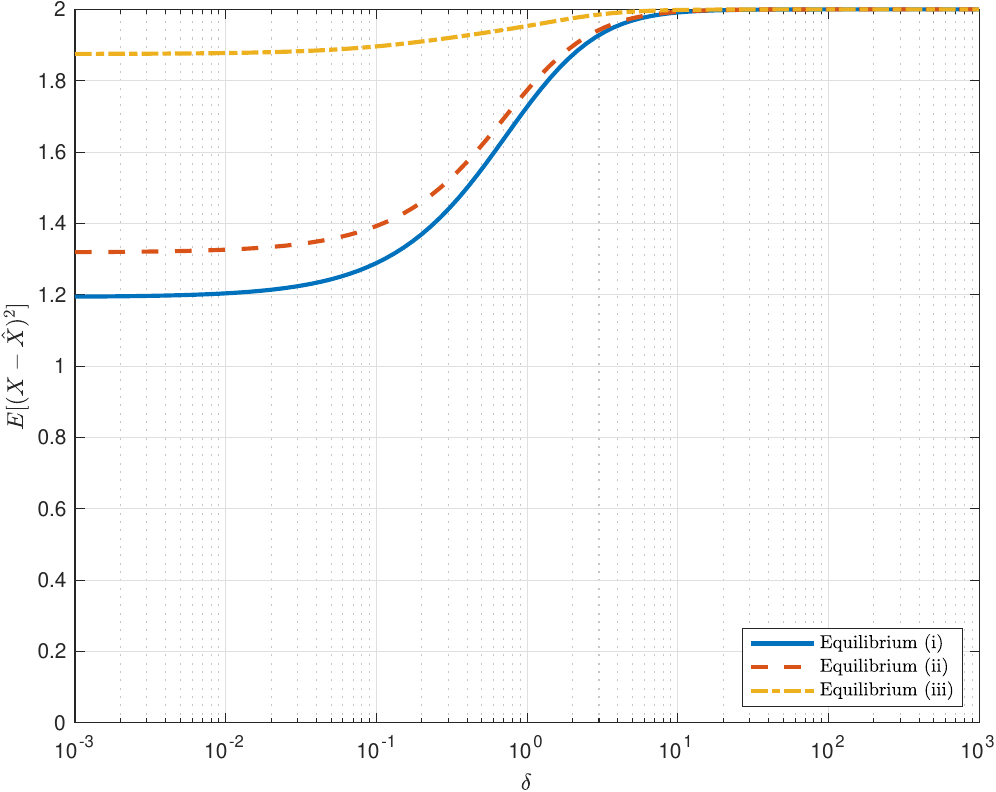}
\caption{Private random variable.}
\label{fig:multidim_private}
\end{subfigure}
\hfill\vspace{5pt}
\begin{subfigure}[t]{0.48\textwidth}
\centering
\includegraphics[width=\textwidth]{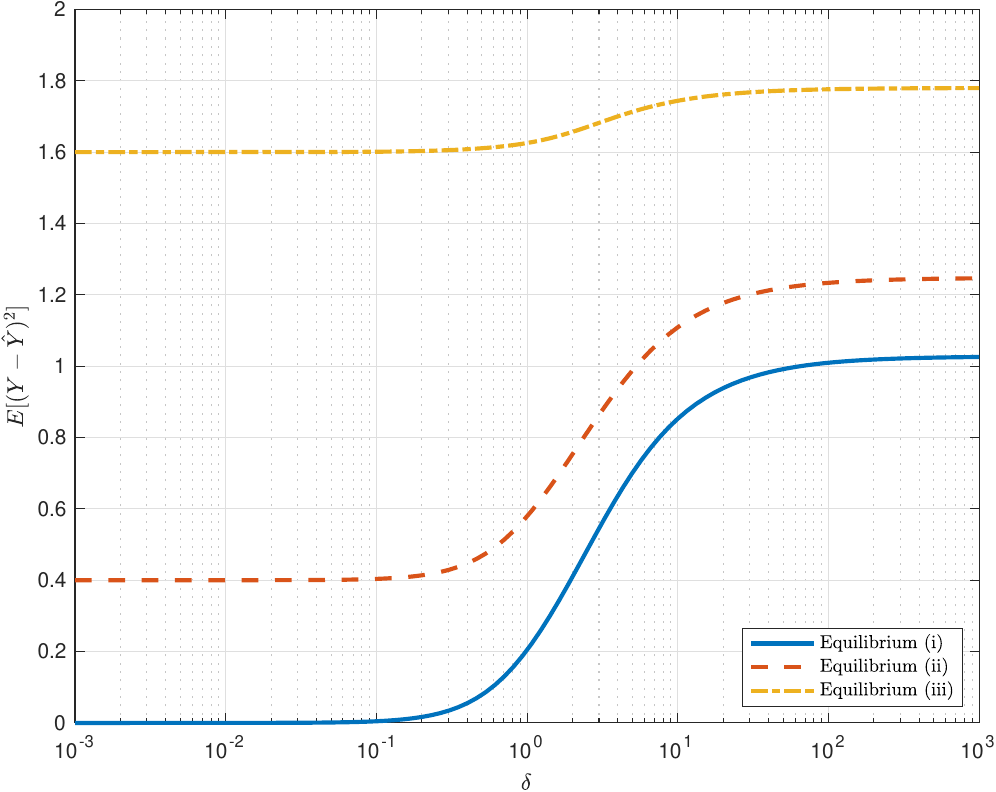}
\caption{Nonprivate random variable.}
\label{fig:multidim_nonprivate}
\end{subfigure}
\hfill\null
\caption{Mean squared errors at the informative equilibria for privacy-signaling game setup with respect to privacy ratio. The equilibrium (i) corresponds to the payoff dominant Nash equilibrium or the Stackelberg equilibrium, and the equilibria (ii) and (iii) correspond to two different Nash equilibria.}
\label{fig:multidim}
\end{figure}
We first illustrate the performance at the equilibria for the privacy-signaling game setup. Since both sources are multidimensional, there exist multiple linear Nash equilibria, which are characterized in Theorem~\ref{thm:NashVector}. Among these Nash equilibria, one of them corresponds to the payoff dominant Nash equilibrium, which also coincides with the Stackelberg equilibrium as stated in Theorem~\ref{thm:StackelbergVector}. We plot the estimation errors at these informative equilibria with respect to the privacy ratio in Fig.~\ref{fig:multidim_private} for the private random variable and in Fig.~\ref{fig:multidim_nonprivate} for the nonprivate random variable. Similar to the scalar source setting, we observe that the estimation performance for both of the random variables degrades as $\delta$ increases since the sender removes more information related to the private random variable and thereby related to the nonprivate random variable. Moreover, the information conveyed at the payoff dominant Nash equilibria contains the information conveyed in other two Nash equilibria. Namely, the sender conveys both $U_1$ and $U_2$ at the payoff dominant Nash equilibria whereas the sender transmits $U_1$ or $U_2$ at the other two Nash equilibria considering the transformed coordinate system defined by \eqref{eq:defT} and \eqref{eq:defUandV}. It is seen that at the equilibrium (iii) in Fig.~\ref{fig:multidim}, the estimation errors do not change significantly with respect to $\delta$ in contrast to that at the equilibrium (ii). This reveals that for the considered setting the tradeoff between privacy and utility is more significant in one direction in the transformed coordinate system. 

\begin{figure}
\centering
\includegraphics[width=\linewidth]{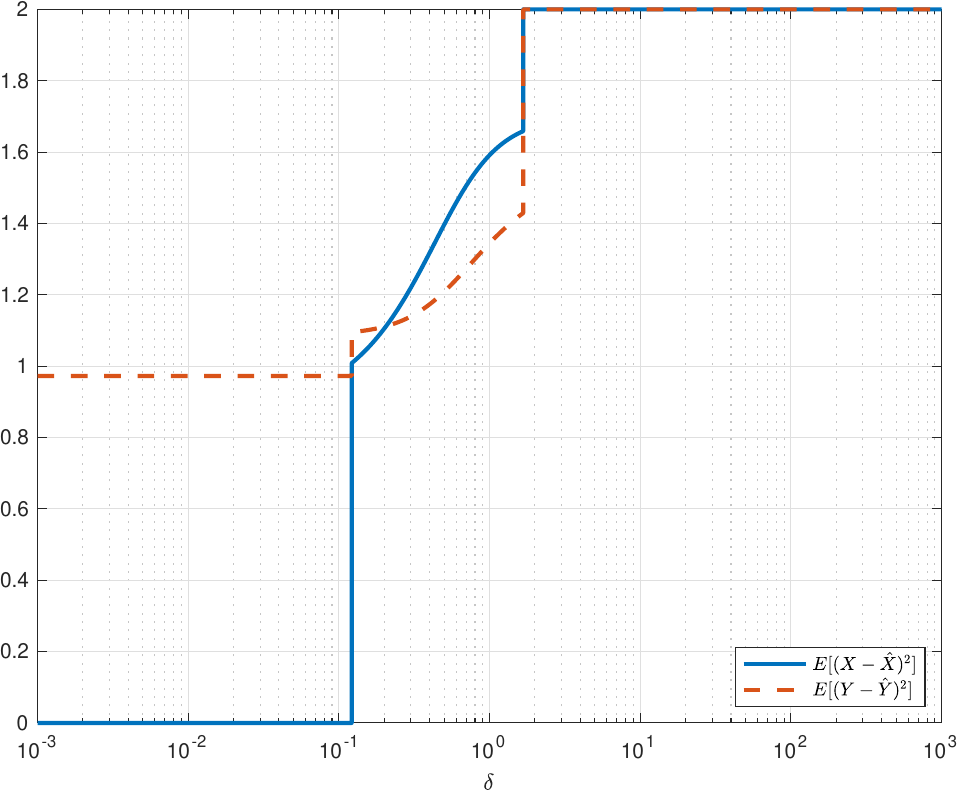}
\caption{Mean squared errors for the MMSE information bottleneck solution with respect to tradeoff parameter $\delta$.}
\label{fig:IBResults}
\end{figure}

\subsection{The MMSE Information Bottleneck Setup}

Finally, we illustrate the performance at the MMSE information bottleneck solution. Fig.~\ref{fig:IBResults} plots the mean squared errors with respect to tradeoff parameter $\delta$. The fully informative scenario with the sender revealing the random variable $\boldsymbol{X}$ completely is obtained for values of $\delta$ smaller than a certain threshold. In contrast, the solution becomes noninformative for values of $\delta$ larger than a certain threshold. When $\delta$ is between these values, the solution becomes informative with the sender applying a certain compression via a linear policy. It is interesting to observe jumps when $\delta$ is equal to the thresholds in Fig.~\ref{fig:IBResults}. In fact, $\delta$ is equal to these thresholds when \eqref{eq:IBEncReduced} has a term with a zero coefficient, i.e., $\lambda_j=0$ for some $j\in\{1,\dots,\xSize\}$. This implies that when $\delta$ is exactly equal to these thresholds, conveying the corresponding random variable $T_j$ in the transformed coordinate system does not affect the sender's cost. As a result, we obtain the optimal solution when the sender transmits $T_j$ as well as when the sender hides $T_j$ completely or partially. Moreover, these thresholds for $\delta$ actually correspond to the values after which the dimension of the encoded message in equilibrium changes.

\section{Conclusion}\label{sec:conc}
A communication setting between a sender with privacy concerns and a receiver has been investigated in a game theoretic framework. The private and nonprivate random variables have been modeled as jointly Gaussian random vectors. It has been proven that a payoff dominant Nash equilibrium is attained by linear policies. It has been shown that these linear policies at the payoff dominant Nash equilibria lead to Stackelberg equilibria as well. These results have been further generalized to the Gaussian noisy channel setting as well as a discrete noiseless channel setting for the special case of scalar sources. We have also provided an estimation theoretic perspective on the information bottleneck problem under the Stackelberg equilibrium concept. We have shown that the Stackelberg equilibria are attained by a set of characterized linear policies. 

\section{Acknowledgments}
The authors would like to thank Prof. Tamer Ba{\c{s}}ar for his detailed and insightful comments and pointing us to \cite{Tamura2018}.

\begin{appendices}
\section{Supporting Results}\label{sec:supporting}

In this appendix, we present supporting results used in the proofs of Theorem~\ref{thm:NashVector} and Theorem~\ref{thm:StackelbergVector}. In the proof of Theorem~\ref{thm:NashVector}, we propose an equivalent formulation by introducing a linear transformation of variables. The following lemma establishes the optimality of the minimum mean squared error estimator at the decoder for a given encoding policy considering the proposed equivalent formulation.

\begin{lemma}\label{lem:MMSEOptimal}
Consider the equivalent formulation illustrated in Fig.~\ref{fig:equivalentBlock} where $\mathcal{T}$ and its inverse are fixed, and the encoder and the decoder select their corresponding policies $\tilde{\gamma}^e(\cdot)$ and $\gamma^{d_{\boldsymbol{T}}}(\cdot)$ arbitrarily. Then, for a fixed encoding function $\tilde{\gamma}^e(\boldsymbol{t})$, the optimal $\gamma^{d_{\boldsymbol{T}}}(\boldsymbol{z})$ that minimizes \eqref{eq:JdEqui} is given by $\Expect[\boldsymbol{T}|\boldsymbol{Z}=\boldsymbol{z}]$. 
\end{lemma}

\begin{IEEEproof}
The result is standard but for completeness we present a short proof. Suppose that $\gamma^{d_{\boldsymbol{T}}}(\boldsymbol{z})= \Expect[\boldsymbol{T}|\boldsymbol{Z}=\boldsymbol{z}] + g(\boldsymbol{z})$. Inserting this expressions into the objective function of the receiver in \eqref{eq:JdEqui}, we get
\begin{align*}
J^d(g) 
&= \Expect[
\left(\boldsymbol{T}-\Expect[\boldsymbol{T}|\boldsymbol{Z}] - g(\boldsymbol{Z})\right)^T
K
\nonumber \\
& \hphantom{=\Expect[\,}
\left(\boldsymbol{T}-
\Expect[\boldsymbol{T}|\boldsymbol{Z}] - g(\boldsymbol{Z})
 \right)] \\
&= \Expect[
\left(\boldsymbol{T}-
\Expect[\boldsymbol{T}|\boldsymbol{Z}]
 \right)^T 
K
\left(\boldsymbol{T}-
\Expect[\boldsymbol{T}|\boldsymbol{Z}]
 \right)
] \nonumber \\
&\hphantom{=}
+ 
\Expect[g(\boldsymbol{Z})^T
K
g(\boldsymbol{Z})]\\
&\geq 
\Expect[
\left(\boldsymbol{T}-
\Expect[\boldsymbol{T}|\boldsymbol{Z}]
 \right)^T 
K
\left(\boldsymbol{T}-
\Expect[\boldsymbol{T}|\boldsymbol{Z}]
 \right)]
\end{align*}
where the inequality follows from $K= Q^T\Sigma Q$ being positive definite. This proves the optimality of the minimum mean squared error estimator in the transformed coordinate system for a given encoding policy.
\end{IEEEproof}

In the following lemma, we show that the sender can only transmit information related to one of the random variables at a Nash equilibrium considering the proposed equivalent formulation illustrated in Fig.\ref{fig:equivalentBlock}.

\begin{lemma}\label{lem:DoNotSendVNash}
Consider the privacy-signaling game problem. At a Nash equilibrium, the sender does not reveal any information related to the linear combinations (of the private and nonprivate random variables) $\boldsymbol{V}$ specified in \eqref{eq:defUandV}.
\end{lemma}

\begin{IEEEproof}
Consider a set of policies where the sender employs an encoding policy $\tilde{\gamma}^e(\boldsymbol{t})=f(\boldsymbol{t})$ which conveys information related to $T_j$ for some $j$ with $\lambda_j<0$. In response to this encoding policy, it is optimal for the receiver to employ the minimum mean squared error estimators of each random variable, as shown in Lemma~\ref{lem:MMSEOptimal}. Denote these estimators for estimating $T_i$ by $g_i(\boldsymbol{z})$ for $i=1,\dots,n$.
Since we assume that the encoding policy $f(\boldsymbol{t})$ conveys information related to $T_j$, the mean squared error for estimating $T_j$ with the corresponding optimal estimator $g_j(\boldsymbol{z})$ is lower than $\sigma_{T_j}^2$, i.e., $\Expect[(T_j-g_j(\boldsymbol{Z}))^2]< \sigma_{T_j}^2$. In response to the decoding policies of $\{g_i(\cdot)\}_{i=1}^n$, the sender can switch to the following policy to improve its objective value. Instead of sending $\boldsymbol{z}=f(\boldsymbol{t})$, the sender can transmit $\boldsymbol{z}=f(\bar{\boldsymbol{t}})$ while keeping the encoding function $f(\cdot)$ the same where $\bar{\boldsymbol{t}} \triangleq [t_1,\dots,t_{j-1},w,t_{j+1},\dots,t_n]^T$ and $w$ is a realization of a random variable that follows the same distribution as $T_j$ and is independent of $\boldsymbol{T}$. In that case, the performance for estimating $T_i$ for $i\neq j$ remains the same since receiving $f(\bar{\boldsymbol{t}})$ or $f(\boldsymbol{t})$ are equivalent for the decoding policy $g_i(\boldsymbol{z})$. However, the performance for estimating $T_j$ degrades as shown in the following:
\begin{align*}
\Expect[(T_j-g_j(\boldsymbol{Z}))^2] 
&= \Expect[T_j^2] - 2 \Expect[T_j g_j(\boldsymbol{Z})] + \Expect[g_j(\boldsymbol{Z})^2] \\
&= \Expect[T_j^2] - 2 \Expect[T_j]\Expect[g_j(\boldsymbol{Z})] + \Expect[g_j(\boldsymbol{Z})^2] \\
&= \Expect[T_j^2] + \Expect[g_j(\boldsymbol{Z})^2] \geq \sigma_{T_j}^2,
\end{align*}
where the second equality is due to the fact that $T_j$ and $\boldsymbol{Z}$ are independent in case $f(\bar{\boldsymbol{T}})$ is transmitted. Since the random variable $T_j$ is chosen such that $\lambda_j<0$ in \eqref{eq:JeEqui}, the sender gains by employing $\boldsymbol{z}=f(\bar{\boldsymbol{t}})$ instead of $\boldsymbol{z}=f(\boldsymbol{t})$. As a result, any encoding policy which yields $\Expect[(T_j-\gamma^{d_{T_j}}(\boldsymbol{Z}))^2]< \sigma_{T_j}^2$ for an index $j$ with $\lambda_j<0$ cannot be a Nash equilibrium since in that case the sender can change its strategy to improve its objective value.

In game theory, when a unilateral change by a decision maker occurs, the perturbed policies may cease to be an equilibrium. However, a subtle aspect of our proof is that, the revised sender policy does not alter the policy of the decoder, therefore the perturbation is still an equilibrium.
\end{IEEEproof}

Similar to the result of Lemma~\ref{lem:DoNotSendVNash} which applies to a Nash equilibrium, the sender is restricted to transmit information related to $\boldsymbol{U}$ at a Stackelberg equilibrium. The following lemma proves this result.

\begin{lemma}\label{lem:DoNotSendVStackelberg}
Consider the privacy-signaling game problem. At a Stackelberg equilibrium, the sender does not reveal any information related to the linear combinations (of the private and nonprivate random variables) $\boldsymbol{V}$ specified in \eqref{eq:defUandV}.
\end{lemma}

\begin{IEEEproof}
We show that any encoding policy which yields $\Expect[(T_j-\gamma^{d_{T_j}}(\boldsymbol{Z}))^2]< \sigma_{T_j}^2$ for an index $j$ with $\lambda_j<0$ cannot be a Stackelberg equilibrium via a similar analysis to that employed in Lemma~\ref{lem:DoNotSendVNash}. Towards that goal, we compare the performance of two scenarios from the perspective of the sender. Recall that in a Stackelberg equilibrium the sender chooses a policy and announces this policy to the receiver and the receiver acts with the knowledge of sender's policy. Denote the encoding policy by $\tilde{\gamma}^e(\boldsymbol{t})=f(\boldsymbol{t})$ in the first scenario. The receiver takes an optimal response to this announced encoding policy. Assume that $\Expect[(T_j-\gamma^{d_{T_j}}(\boldsymbol{Z}))^2]< \sigma_{T_j}^2$ with the corresponding set of policies. In the second scenario, suppose that the encoder chooses the same policy as before with the exception that the sender replaces the realization $T_j=t_j$ by an independent noise following the same distribution as $T_j$. Namely, the sender uses $f(\bar{\boldsymbol{t}})$ where $\bar{\boldsymbol{t}} = [t_1,\dots,t_{j-1},w,t_{j+1},\dots,t_n]^T$ and $w$ is a realization of a random variable that follows the same distribution as $T_j$ and is independent of $\boldsymbol{T}$. As the sender announces its strategy, the optimal response of the receiver for the random variable $T_j$ becomes $\gamma^{d_{T_j}}(\boldsymbol{z})=\Expect[T_j|\boldsymbol{Z}=\boldsymbol{z}]=\Expect[T_j]=0$ due to the independence of $T_j$ and $\boldsymbol{Z}$ in this scenario. Therefore, we get $\Expect[(T_j-\gamma^{d_{T_j}}(\boldsymbol{Z}))^2]=\sigma_{T_j}^2$ in this case. Notice that the mean squared error performance in estimating $T_i$ for $i\neq j$ is the same for both scenarios. As a result, the second scenario yields better performance for the sender. Thus, transmitting information related to $T_j$ with $\lambda_j<0$ in \eqref{eq:JeEqui} is not desirable for the sender.
\end{IEEEproof}

\end{appendices}

\bibliographystyle{IEEEtran}
\bibliography{PrivacyGames}

\begin{thebibliography}{10}
\providecommand{\url}[1]{#1}
\csname url@samestyle\endcsname
\providecommand{\newblock}{\relax}
\providecommand{\bibinfo}[2]{#2}
\providecommand{\BIBentrySTDinterwordspacing}{\spaceskip=0pt\relax}
\providecommand{\BIBentryALTinterwordstretchfactor}{4}
\providecommand{\BIBentryALTinterwordspacing}{\spaceskip=\fontdimen2\font plus
\BIBentryALTinterwordstretchfactor\fontdimen3\font minus
  \fontdimen4\font\relax}
\providecommand{\BIBforeignlanguage}[2]{{%
\expandafter\ifx\csname l@#1\endcsname\relax
\typeout{** WARNING: IEEEtran.bst: No hyphenation pattern has been}%
\typeout{** loaded for the language `#1'. Using the pattern for}%
\typeout{** the default language instead.}%
\else
\language=\csname l@#1\endcsname
\fi
#2}}
\providecommand{\BIBdecl}{\relax}
\BIBdecl

\bibitem{ISITVersion}
E.~Kaz{\i}kl{\i}, S.~Gezici, and S.~Y\"uksel, ``Quadratic privacy-signaling
  games and payoff dominant equilibria,'' in \emph{IEEE International Symposium
  on Information Theory (ISIT)}, 2020.

\bibitem{McDaniel2009}
P.~McDaniel and S.~McLaughlin, ``Security and privacy challenges in the smart
  grid,'' \emph{{IEEE} Security Privacy}, vol.~7, no.~3, pp. 75--77, May 2009.

\bibitem{PrivacySmartMeteringSurvey}
S.~Finster and I.~Baumgart, ``Privacy-aware smart metering: A survey,''
  \emph{{IEEE} Communications Surveys Tutorials}, vol.~17, no.~2, pp.
  1088--1101, Secondquarter 2015.

\bibitem{HanACC2016}
S.~{Han}, U.~{Topcu}, and G.~J. {Pappas}, ``Event-based information-theoretic
  privacy: A case study of smart meters,'' in \emph{American Control Conference
  (ACC)}, 2016, pp. 2074--2079.

\bibitem{CompetiviePrivacy2011}
L.~Sankar, S.~Kar, R.~Tandon, and H.~V. Poor, ``Competitive privacy in the
  smart grid: An information-theoretic approach,'' in \emph{{IEEE}
  International Conference on Smart Grid Communications}, Oct. 2011, pp.
  220--225.

\bibitem{YaoAllerton2013}
J.~{Yao} and P.~{Venkitasubramaniam}, ``On the privacy-cost tradeoff of an
  in-home power storage mechanism,'' in \emph{Annual Allerton Conference on
  Communication, Control, and Computing (Allerton)}, 2013, pp. 115--122.

\bibitem{CrowdSensingWirelessComm2015}
D.~He, S.~Chan, and M.~Guizani, ``User privacy and data trustworthiness in
  mobile crowd sensing,'' \emph{{IEEE} Wireless Communications}, vol.~22,
  no.~1, pp. 28--34, Feb. 2015.

\bibitem{PrivacyCrowdSensingIoT2016}
S.~Gisdakis, T.~Giannetsos, and P.~Papadimitratos, ``Security, privacy, and
  incentive provision for mobile crowd sensing systems,'' \emph{{IEEE} Internet
  of Things Journal}, vol.~3, no.~5, pp. 839--853, Oct. 2016.

\bibitem{Yamamoto1983}
H.~{Yamamoto}, ``A source coding problem for sources with additional outputs to
  keep secret from the receiver or wiretappers (corresp.),'' \emph{IEEE
  Transactions on Information Theory}, vol.~29, no.~6, pp. 918--923, Nov. 1983.

\bibitem{EstEffUndPriConstr2019}
S.~{Asoodeh}, M.~{Diaz}, F.~{Alajaji}, and T.~{Linder}, ``Estimation efficiency
  under privacy constraints,'' \emph{{IEEE} Transactions on Information
  Theory}, vol.~65, no.~3, pp. 1512--1534, March 2019.

\bibitem{WangIT2019}
H.~{Wang}, L.~{Vo}, F.~P. {Calmon}, M.~{Médard}, K.~R. {Duffy}, and
  M.~{Varia}, ``Privacy with estimation guarantees,'' \emph{IEEE Transactions
  on Information Theory}, vol.~65, no.~12, pp. 8025--8042, 2019.

\bibitem{PadakandlaIT2020}
A.~{Padakandla}, P.~R. {Kumar}, and W.~{Szpankowski}, ``The trade-off between
  privacy and fidelity via {E}hrhart theory,'' \emph{IEEE Transactions on
  Information Theory}, vol.~66, no.~4, pp. 2549--2569, 2020.

\bibitem{DiazIT2020}
M.~{Diaz}, H.~{Wang}, F.~P. {Calmon}, and L.~{Sankar}, ``On the robustness of
  information-theoretic privacy measures and mechanisms,'' \emph{IEEE
  Transactions on Information Theory}, vol.~66, no.~4, pp. 1949--1978, 2020.

\bibitem{UtilityvsPrivacy2013}
L.~{Sankar}, S.~R. {Rajagopalan}, and H.~V. {Poor}, ``Utility-privacy tradeoffs
  in databases: An information-theoretic approach,'' \emph{IEEE Transactions on
  Information Forensics and Security}, vol.~8, no.~6, pp. 838--852, June 2013.

\bibitem{UtilityvsPrivacy2018}
B.~{Rassouli} and D.~{Gündüz}, ``Optimal utility-privacy trade-off with total
  variation distance as a privacy measure,'' \emph{IEEE Transactions on
  Information Forensics and Security}, vol.~15, pp. 594--603, 2020.

\bibitem{SreekumarISIT2020}
S.~{Sreekumar} and D.~{Gündüz}, ``Optimal privacy-utility trade-off under a
  rate constraint,'' in \emph{IEEE International Symposium on Information
  Theory (ISIT)}, 2019, pp. 2159--2163.

\bibitem{Calmon2012}
F.~P. {Calmon} and N.~{Fawaz}, ``Privacy against statistical inference,'' in
  \emph{Annual Allerton Conference on Communication, Control, and Computing
  (Allerton)}, Oct. 2012, pp. 1401--1408.

\bibitem{LuAutomatica2020}
Y.~Lu and M.~Zhu, ``On privacy preserving data release of linear dynamic
  networks,'' \emph{Automatica}, vol. 115, p. 108839, 2020.

\bibitem{DiffPrivateFilteringTAC2014}
J.~{Le Ny} and G.~J. {Pappas}, ``Differentially private filtering,'' \emph{IEEE
  Transactions on Automatic Control}, vol.~59, no.~2, pp. 341--354, 2014.

\bibitem{PrivacyConsensusTAC2017}
Y.~{Mo} and R.~M. {Murray}, ``Privacy preserving average consensus,''
  \emph{IEEE Transactions on Automatic Control}, vol.~62, no.~2, pp. 753--765,
  2017.

\bibitem{Nekouei2019}
E.~Nekouei, T.~Tanaka, M.~Skoglund, and K.~H. Johansson,
  ``Information-theoretic approaches to privacy in estimation and control,''
  \emph{Annual Reviews in Control}, vol.~47, pp. 412 -- 422, 2019.

\bibitem{WangTCNS2017}
Y.~{Wang}, Z.~{Huang}, S.~{Mitra}, and G.~E. {Dullerud}, ``Differential privacy
  in linear distributed control systems: Entropy minimizing mechanisms and
  performance tradeoffs,'' \emph{IEEE Transactions on Control of Network
  Systems}, vol.~4, no.~1, pp. 118--130, 2017.

\bibitem{TBasarBook}
T.~Ba{\c{s}}ar and G.~J. Olsder, \emph{Dynamic Noncooperative Game
  Theory}.\hskip 1em plus 0.5em minus 0.4em\relax Philadelphia, PA: SIAM
  Classics in Applied Mathematics, 1999.

\bibitem{kamenica2011bayesian}
E.~Kamenica and M.~Gentzkow, ``Bayesian persuasion,'' \emph{American Economic
  Review}, vol. 101, no.~6, pp. 2590--2615, Oct. 2011.

\bibitem{tishby1999information}
N.~Tishby, F.~C. Pereira, and W.~Bialek, ``The information bottleneck method,''
  in \emph{Annual Allerton Conference on Communication, Control, and Computing
  (Allerton)}, 1999, pp. 368--377.

\bibitem{CrawfordSobel}
V.~P. Crawford and J.~Sobel, ``Strategic information transmission,''
  \emph{Econometrica}, vol.~50, no.~6, pp. 1431--1451, 1982.

\bibitem{Tamura2018}
\BIBentryALTinterwordspacing
W.~Tamura, ``Bayesian persuasion with quadratic preferences.'' [Online].
  Available: \url{https://ssrn.com/abstract=1987877}
\BIBentrySTDinterwordspacing

\bibitem{Ekyol2017ProcIEEE}
E.~Akyol, C.~Langbort, and T.~Ba{\c{s}}ar, ``Information-theoretic approach to
  strategic communication as a hierarchical game,'' \emph{Proceedings of the
  IEEE}, vol. 105, no.~2, pp. 205--218, Feb. 2017.

\bibitem{Saritas2017QuadraticEquilibria}
S.~Sar{\i}ta{\c{s}}, S.~Y{\"u}ksel, and S.~Gezici, ``Quadratic
  multi-dimensional signaling games and affine equilibria,'' \emph{IEEE
  Transactions on Automatic Control}, vol.~62, no.~2, pp. 605--619, Feb. 2017.

\bibitem{EstStrategicSensorsFarokhi2017}
F.~Farokhi, A.~M.~H. Teixeira, and C.~Langbort, ``Estimation with strategic
  sensors,'' \emph{IEEE Transactions on Automatic Control}, vol.~62, no.~2, pp.
  724--739, Feb. 2017.

\bibitem{SubjectiveBiasesSITBasar2018}
V.~S.~S. Nadendla, C.~Langbort, and T.~Ba{\c{s}}ar, ``Effects of subjective
  biases on strategic information transmission,'' \emph{IEEE Transactions on
  Communications}, vol.~66, no.~12, pp. 6040--6049, Dec. 2018.

\bibitem{treust2017persuasion}
M.~L. Treust and T.~Tomala, ``Persuasion with limited communication capacity,''
  \emph{Journal of Economic Theory}, vol. 184, p. 104940, 2019.

\bibitem{treust2018persuasion}
\BIBentryALTinterwordspacing
------, ``Information-theoretic limits of strategic communication,'' 2018.
  [Online]. Available: \url{http://arxiv.org/abs/1807.05147}
\BIBentrySTDinterwordspacing

\bibitem{SaritasAutomatica2020}
S.~Sar{\i}ta{\c{s}}, S.~Y{\"u}ksel, and S.~Gezici, ``Dynamic signaling games
  with quadratic criteria under {N}ash and {S}tackelberg equilibria,''
  \emph{Automatica}, vol. 115, p. 108883, 2020.

\bibitem{SaritasISIT2019}
S.~Sar{\i}ta{\c{s}}, P.~Furrer, S.~Gezici, T.~Linder, and S.~Y{\"u}ksel, ``On
  the number of bins in equilibria for signaling games,'' in \emph{IEEE
  International Symposium on Information Theory (ISIT)}, 2019, pp. 972--976.

\bibitem{SayinAutomatica2019}
M.~O. Say{\i}n, E.~Akyol, and T.~Ba{\c{s}}ar, ``Hierarchical multistage
  {G}aussian signaling games in noncooperative communication and control
  systems,'' \emph{Automatica}, vol. 107, pp. 9 -- 20, 2019.

\bibitem{sayin2019optimal}
M.~O. Say{\i}n and T.~Ba{\c{s}}ar, ``Bayesian persuasion with state-dependent
  quadratic cost measures,'' \emph{IEEE Transactions on Automatic Control},
  vol.~67, no.~3, pp. 1241--1252, 2022.

\bibitem{MultiCheapArxiv}
E.~Kaz{\i}kl{\i}, S.~Gezici, and S.~Y\"uksel, ``Signaling games in higher
  dimensions: Geometric properties of equilibrium solutions,'' \emph{arXiv
  preprint: 2108.05240}, 2021.

\bibitem{Farokhi2015QuadraticGames}
F.~Farokhi, H.~Sandberg, I.~Shames, and M.~Cantoni, ``Quadratic {G}aussian
  privacy games,'' in \emph{IEEE Conference on Decision and Control (CDC)},
  Dec. 2015, pp. 4505--4510.

\bibitem{Akyol2015PrivacyProcessing}
E.~Akyol, C.~Langbort, and T.~Ba{\c{s}}ar, ``Privacy constrained information
  processing,'' in \emph{IEEE Conference on Decision and Control (CDC)}, Dec.
  2015, pp. 4511--4516.

\bibitem{Farokhi2016PrivacyCommunication}
F.~Farokhi and G.~Nair, ``Privacy-constrained communication,''
  \emph{IFAC-PapersOnLine}, vol.~49, no.~22, pp. 43--48, Jan 2016.

\bibitem{EAkyolITW2015}
E.~Akyol, C.~Langbort, and T.~Ba{\c{s}}ar, ``Strategic compression and
  transmission of information,'' in \emph{IEEE Information Theory Workshop -
  Fall}, Oct. 2015, pp. 219--223.

\bibitem{DifPrivacy2008}
C.~Dwork, ``Differential privacy: A survey of results,'' in \emph{Theory and
  Applications of Models of Computation}, M.~Agrawal, D.~Du, Z.~Duan, and
  A.~Li, Eds.\hskip 1em plus 0.5em minus 0.4em\relax Berlin, Heidelberg:
  Springer, 2008, pp. 1--19.

\bibitem{AlgoDifPrivacy2014}
C.~Dwork and A.~Roth, ``The algorithmic foundations of differential privacy,''
  \emph{Foundations and Trends in Theoretical Computer Science}, vol.~9, no.
  3–4, pp. 211--407, 2014.

\bibitem{Calmon2015}
F.~P. {Calmon}, A.~{Makhdoumi}, and M.~{Médard}, ``Fundamental limits of
  perfect privacy,'' in \emph{IEEE International Symposium on Information
  Theory (ISIT)}, June 2015, pp. 1796--1800.

\bibitem{IBtoPrivacyFunnel2014}
A.~{Makhdoumi}, S.~{Salamatian}, N.~{Fawaz}, and M.~{Medard}, ``From the
  information bottleneck to the privacy funnel,'' in \emph{IEEE Information
  Theory Workshop (ITW)}, Nov. 2014, pp. 501--505.

\bibitem{HanTAC2017}
S.~{Han}, U.~{Topcu}, and G.~J. {Pappas}, ``Differentially private distributed
  constrained optimization,'' \emph{IEEE Transactions on Automatic Control},
  vol.~62, no.~1, pp. 50--64, 2017.

\bibitem{jeromeBook}
J.~Le~Ny, \emph{Differential Privacy for Dynamic Data}.\hskip 1em plus 0.5em
  minus 0.4em\relax Springer, 2020.

\bibitem{TishbyITW2015}
N.~{Tishby} and N.~{Zaslavsky}, ``Deep learning and the information bottleneck
  principle,'' in \emph{IEEE Information Theory Workshop (ITW)}, 2015, pp.
  1--5.

\bibitem{HarremoesISIT2007}
P.~Harremo{\"e}s and N.~Tishby, ``The information bottleneck revisited or how
  to choose a good distortion measure,'' in \emph{IEEE International Symposium
  on Information Theory}, 2007, pp. 566--570.

\bibitem{Gilad-Bachrach2003}
R.~Gilad-Bachrach, A.~Navot, and N.~Tishby, ``An information theoretic tradeoff
  between complexity and accuracy,'' in \emph{Learning Theory and Kernel
  Machines}.\hskip 1em plus 0.5em minus 0.4em\relax Springer, 2003, pp.
  595--609.

\bibitem{Dhillon2003}
I.~S. Dhillon, S.~Mallela, and D.~S. Modha, ``Information-theoretic
  co-clustering,'' in \emph{Proceedings of the Ninth ACM SIGKDD International
  Conference on Knowledge Discovery and Data Mining}.\hskip 1em plus 0.5em
  minus 0.4em\relax Association for Computing Machinery, 2003, p. 89–98.

\bibitem{VeraISIT2018}
M.~{Vera}, P.~{Piantanida}, and L.~R. {Vega}, ``The role of the information
  bottleneck in representation learning,'' in \emph{IEEE International
  Symposium on Information Theory (ISIT)}, 2018, pp. 1580--1584.

\bibitem{VeraIT2019}
M.~{Vera}, L.~{Rey Vega}, and P.~{Piantanida}, ``Collaborative information
  bottleneck,'' \emph{IEEE Transactions on Information Theory}, vol.~65, no.~2,
  pp. 787--815, 2019.

\bibitem{GoldfeldJSAIT2020}
Z.~{Goldfeld} and Y.~{Polyanskiy}, ``The information bottleneck problem and its
  applications in machine learning,'' \emph{IEEE Journal on Selected Areas in
  Information Theory}, vol.~1, no.~1, pp. 19--38, 2020.

\bibitem{HsuISIT2018}
H.~{Hsu}, S.~{Asoodeh}, S.~{Salamatian}, and F.~P. {Calmon}, ``Generalizing
  bottleneck problems,'' in \emph{IEEE International Symposium on Information
  Theory (ISIT)}, 2018, pp. 531--535.

\bibitem{ZaidiEntropy2020}
A.~Zaidi, I.~Estella-Aguerri, and S.~Shamai~(Shitz), ``On the information
  bottleneck problems: Models, connections, applications and information
  theoretic views,'' \emph{Entropy}, vol.~22, no.~2, p. 151, Jan 2020.

\bibitem{WitsenhausenIT1975}
H.~{Witsenhausen} and A.~{Wyner}, ``A conditional entropy bound for a pair of
  discrete random variables,'' \emph{IEEE Transactions on Information Theory},
  vol.~21, no.~5, pp. 493--501, 1975.

\bibitem{chechik2005IB}
G.~Chechik, A.~Globerson, N.~Tishby, and Y.~Weiss, ``Information bottleneck for
  {G}aussian variables,'' \emph{Journal of Machine Learning Research}, vol.~6,
  no. Jan, pp. 165--188, 2005.

\bibitem{HarsanyiSeltenBook1988}
J.~C. Harsanyi and R.~Selten, \emph{A General Theory of Equilibrium Selection
  in Games}.\hskip 1em plus 0.5em minus 0.4em\relax Cambridge, Massachusets:
  MIT Press, 1988.

\bibitem{horn2013matrixAnalysis}
R.~A. Horn and C.~R. Johnson, \emph{Matrix Analysis}.\hskip 1em plus 0.5em
  minus 0.4em\relax New York: Cambridge University Press, 2013.

\bibitem{PoorBook}
H.~V. Poor, \emph{An Introduction to Signal Detection and Estimation}.\hskip
  1em plus 0.5em minus 0.4em\relax New York: Springer-Verlag, 1994.

\bibitem{IBOptimality}
A.~Globerson and N.~Tishby, ``On the optimality of {G}aussian information
  bottleneck curve,'' Hebrew Univ, Jerusalem, Israel, Tech. Rep., 2004.

\bibitem{YukselBook2013}
S.~Y{\"u}ksel and T.~Ba{\c{s}}ar, \emph{Stochastic Networked Control Systems:
  Stabilization and Optimization under Information Constraints}.\hskip 1em plus
  0.5em minus 0.4em\relax Boston, MA: Birkhauser, 2013.

\bibitem{GrayNeuhoffQuantization}
R.~M. {Gray} and D.~L. {Neuhoff}, ``Quantization,'' \emph{IEEE Transactions on
  Information Theory}, vol.~44, no.~6, pp. 2325--2383, Oct. 1998.

\end{thebibliography}

\end{document}